\documentclass[amsmath,amssymb,aps,prd,twocolumn,superscriptaddress,showpacs]{revtex4}
\usepackage[dvips]{graphicx}
\usepackage{times}
\newcommand{\nn}{\nonumber}

\newcommand{\VEV}[1]{\left\langle{#1}\right\rangle}
\newcommand{\chibar}{{\bar{\chi}}}
\newcommand{\Psfig}[2]{\includegraphics[width=#1]{Figs/#2}}
\newcommand{\Feff}[1]{{\cal F}_\mathrm{#1}}
\newcommand{\Vq}{{\cal V}_{q}}

\newcommand{\Od}{{\cal O}}
\newcommand{\Vp}{V^{+}}
\newcommand{\Vm}{V^{-}}
\newcommand{\bsig}{b_\sigma}

\newcommand{\ssB}{{\scriptscriptstyle B}}
\newcommand{\rhoB}{\rho_{\scriptscriptstyle B}}

\newcommand{\wt}{{\omega_\tau}}

\newcommand{\Disp}[1]{${\displaystyle #1}$}
\begin{document}
%\preprint{}

\title{Phase diagram evolution at finite coupling in strong coupling lattice QCD}

\author{Kohtaroh Miura}
\email[]{miura@yukawa.kyoto-u.ac.jp}
%\thanks{}
\affiliation{Yukawa Institute for Theoretical Physics, Kyoto University,
Kyoto 606-8502, Japan}
%%%%%
\author{Takashi Z. Nakano}
\affiliation{Department of Physics, Faculty of Science, Kyoto University,
Kyoto 606-8502, Japan}
%%%%%
%%%%%
\author{Akira Ohnishi}
\affiliation{Yukawa Institute for Theoretical Physics, Kyoto University,
Kyoto 606-8502, Japan}
%%%%%
\author{Noboru Kawamoto}
\affiliation{Department of Physics, Faculty of Science, Hokkaido University,
Sapporo 060-0810, Japan}

\date{\today}
\pacs{11.15.Me, 12.38.Gc, 11.10.Wx, 25.75.Nq}

\begin{abstract}
We investigate the chiral phase transition
in the strong coupling lattice QCD
at finite temperature and density with finite coupling effects.
We adopt one species of staggered fermion,
and develop an analytic formulation based 
on strong coupling and cluster expansions.
We derive the effective potential as
a function of two order parameters, 
the chiral condensate $\sigma$
and the quark number density $\rho_q$,
in a self-consistent treatment of
the next-to-leading order (NLO) effective action terms.
NLO contributions lead to modifications of
quark mass, chemical potential and
the quark wave function renormalization factor.
While the ratio $\mu_c(T=0)/T_c(\mu=0)$
is too small in the strong coupling limit, 
it is found to increase as $\beta=2N_c/g^2$ increases.
The critical point is found to move in the lower $T$ direction
as $\beta$ increases.
Since the vector interaction induced by $\rho_q$
is shown to grow as $\beta$,
the present trend is consistent with the results
in Nambu-Jona-Lasinio models.
The interplay between two order parameters leads to
the existence of partially chiral restored matter,
where effective chemical potential 
is automatically adjusted to the quark excitation energy.
\end{abstract}
\maketitle
%%%%%%%%%%%%%%%%%%%%%%%%%%%%%%%%%%%%%%
\section{Introduction}
Exploring the chiral phase transition
and its phase diagram
in Quantum Chromodynamics (QCD)
is one of the most challenging problems
in quark hadron physics.
The chiral phase transition
may really happen in compact astrophysical phenomena
such as the early universe and compact stars,
and can be investigated 
in heavy-ion collision experiments.
The large magnitude of the elliptic flow parameter
observed in the relativistic heavy-ion collider (RHIC) experiments
indicates the formation of strongly interacting 
quark-gluon plasma~\cite{Muller:2006ee,STAR,PHENIX,PHOBOS}
at high temperature.
The future experiments at FAIR and in low energy programs at RHIC
are expected to provide new discoveries
in the phase diagram investigations.

The most rigorous and reliable framework
to investigate the QCD phase transition
would be the lattice QCD Monte-Carlo (MC) simulations.
In the high temperature and low density region,
the lattice MC can provide the quantitative predictions,
and the critical temperature 
is estimated to be $T_c\simeq (160-190)~\mathrm{MeV}$~\cite{LQCD_T,LQCD_Tmu}.
In comparison, the lattice MC simulations
do not work well in the high baryon density region
because of the notorious sign problem of the Dirac determinant.
Many ideas have been proposed to overcome this problem~\cite{LQCD_mu},
for example,
the Taylor expansion around $\mu=0$~\cite{Taylor},
analytic continuation~\cite{AC,D'Elia:2002gd},
canonical ensemble method~\cite{Canonical},
improved reweighting method~\cite{FodorKatz},
and the density of states method~\cite{Fodor:2007vv}.
It has become possible to access the
relatively small density region
$\mu/T \lesssim 1.0$ \cite{LQCD_Tmu,LQCD_mu}.
One of the interesting objects is
the critical end point (CEP)~\cite{CEP}.
Recent works indicate that
CEP may locate in the region $\mu/T \geq 1.0$~\cite{BSCorlab,Aoki:2008rt,Ejiri:2007ga},
while de Forcrand and Philipsen suggest
that CEP might not exist~\cite{deForcrand:2008vr}.
The larger chemical potential region is still under debate, 
and alternative methods are necessary to
reveal the whole structure of the phase diagram.

In the phase diagram investigation,
the strong coupling lattice QCD (SC-LQCD), 
the lattice QCD formulation
based on the expansion of the action in the power series
of the inverse bare coupling squared ($1/g^2$),
is an interesting approach,
since the sign problem can be weakened or avoided.
The SC-LQCD was applied first to the confinement study 
in pure Yang-Mills theories~\cite{PureGlueRevSCLQCD,Drouffe:1983fv,Wilson:1974sk}.
Wilson suggested the confinement mechanism in an analytic study
of the strong coupling limit (SCL) of lattice QCD~\cite{Wilson:1974sk}.
Creutz showed that the $\beta=2N_c/g^2$ dependence
of the lattice spacing $a$ in the MC simulation
smoothly connects the strong coupling behavior
and the continuum spacetime scaling behavior~\cite{Creutz}.
By using the character expansion technique,
M\"unster demonstrated that
the pure Yang-Mills SC-LQCD with high order corrections
could explain the above MC results~\cite{Munster:1980iv}.
The scaling behavior in MC simulations indicates
that the confinement is actually realized 
in the continuum spacetime,
and the success of SC-LQCD suggests that
the scaling region would be accessible in SC-LQCD
within the conversion radius, which is shown to be finite
in pure Yang-Mills theories~\cite{convergence}.
We may expect that the scaling and convergent properties
are also kept with fermions.
Then it would be possible that the SC-LQCD could provide useful results
on the phase diagram in the whole region of the $T-\mu$ plane.

The SC-LQCD with fermions has a long history
of chiral symmetry studies for more than twenty
years~\cite{Kawamoto:1981hw,Hoek:1981uv,Kawamoto:1982gw,Smit:1980nf,KlubergStern:1981wz,KlubergStern:1982bs,KlubergStern:1983dg,Jolicoeur:1983tz,Ichinose,DKS,DHK1985,Faldt1986,Dagotto,Karsch:1988zx,BDP,BKR,Bilic:1995tq,Nishida:2003uj,Fukushima:2003vi,Nishida:2003fb,Azcoiti:2003eb,Kawamoto:2005mq,Lat08_Miura,QY,Mmass,Ohnishi:2007eg,Ohnishi:2007fb,Bringoltz:2006pz,deForcrand:2009dh,Levi:1996rn,Umino,Fang:2001ry,Luo:2004mc,Brower:1999ak,Levkova:2004xw,IchinoseNagao,XQLuo},
and many theoretical tools have been developed;
the large dimensional or $1/d$ ($d=$ spatial or spacetime dimension) 
expansion~\cite{KlubergStern:1982bs},
the finite temperature treatments
in the Polyakov gauge \cite{DKS} 
and in the temporal gauge \cite{Faldt1986},
the finite quark chemical potential effect~\cite{DHK1985}
with the help of the lattice chemical potential \cite{Hasenfratz:1983ba}.
The analytic expression of the SCL effective potential has been derived
at finite $T$~\cite{DKS,Faldt1986} or at finite $\mu$~\cite{DHK1985}.
In 1990's,
phase diagram studies met some successes~\cite{BDP,BKR,Bilic:1995tq}
based on the SC-LQCD effective action~\cite{Faldt1986}.
We also find several works on the Polyakov loop~\cite{V_Ploop}
based on SC-LQCD,
and the functional form of the effective potential in SC-LQCD
has provided basic ingredients
in the SC-LQCD related models~\cite{Ilgenfritz:1984ff,Gocksch:1984yk,SCL_PLoop}
and the Nambu-Jona-Lasinio model with Polyakov loop (PNJL)~\cite{Fukushima:2003fw}.

Based on these successes,
the SC-LQCD is recently revisited and expected to provide
an instructive guide to QCD under extreme conditions.
The pure Yang-Mills SC-LQCD~\cite{Munster:1980iv} is extended
to finite $T$~\cite{Langelage:2008dj},
and the shear viscosity below the deconfinement transition temperature
is also studied~\cite{Jakovac:2008ft}.
In color SU$(2)$ QCD, the interplay between the diquark condensate $\Delta$ 
and the chiral condensate $\sigma$ is investigated
at finite $T$ and $\mu$~\cite{Nishida:2003uj}.
The diquark effect is also investigated for color SU$(3)$
at zero temperature~\cite{Azcoiti:2003eb}.
We find remarkable developments in the SU($N_c=3$) phase diagram
investigations~\cite{Fukushima:2003vi,Nishida:2003fb,Kawamoto:2005mq,deForcrand:2009dh},
where a ``na\"ive'' structure
of the phase diagram with first and second transition lines
separated by a tri-critical point (TCP) is obtained
in the strong coupling limit with zero quark masses.
%Recently, these analytic results are 
%qualitatively reproduced by
%``a strong coupling limit of lattice MC simulation''~\cite{deForcrand:2009dh}
%based on a Monomer-Dimer-Polymer formulation~\cite{Karsch:1988zx}.
With non-zero quark masses,
TCP becomes a critical end point (CEP),
whose discovery is one of the physics goals
in low-energy programs at RHIC.

In order to discuss
the chiral symmetry on the lattice,
the SC-LQCD has been developed in several fermion formalisms.
We find some pioneering works
based on the staggered~\cite{Kawamoto:1981hw,Hoek:1981uv}, 
the Wilson~\cite{Smit:1980nf,Kawamoto:1981hw}
and the na\"ive~\cite{KlubergStern:1981wz} fermions.
The domain-wall \cite{Kaplan:1992bt} 
and the overlap \cite{Neuberger:1998wv} fermion
provide modern formulation of the lattice chiral symmetry,
and some SC-LQCD based investigations are found in
\cite{Brower:1999ak,Levkova:2004xw} (domain-wall) and
\cite{IchinoseNagao,XQLuo} (overlap).
In the present work,
we adopt one species ($n_f=1$) of (unrooted) staggered fermion.
Its simple realization of the chiral symmetry on the
lattice~\cite{Susskind:1976jm,Sharatchandra:1981si,Kawamoto:1981hw}
is useful to develop analytic formulations.
It has been theoretically suggested \cite{KlubergStern:1983dg,Golt_Smit} and 
numerically established \cite{stagg} that
the unrooted staggered QCD is equivalent to
the four flavor ($N_f=4$) QCD with degenerate masses
in the continuum limit.

In this paper,
we investigate the phase diagram evolution with finite coupling effects.
We employ one species of (unrooted) staggered fermion,
and take account of the next-to-leading order (NLO, ${\cal O}(1/g^2)$) terms
in the strong coupling expansion.
We concentrate on the leading order
of the large dimensional ($1/d$)
expansion~\cite{KlubergStern:1982bs} for simplicity.
The gluon field is evaluated in the Polyakov gauge~\cite{DKS}
with respect for the finite temperature $T$ effects,
and the finite density effects are introduced
via the quark chemical potential $\mu$ on the lattice~\cite{Hasenfratz:1983ba}.
In these setups, we derive an analytic expression of the effective potential
in the mean field approximation.
In particular,
the following points are newly developed.
Firstly
we introduce the NLO effective action terms
through the systematic cluster expansion.
Secondly we evaluate the NLO effective action
by using a recently developed extended Hubbard-Stratonovich (EHS)
transformation~\cite{QY,Lat08_Miura}.
As a result, several auxiliary fields including the chiral condensate
$\sigma$ are introduced on the same footing,
and the NLO effects are self-consistently evaluated.
In particular,
we find that the quark number density naturally appears as an order parameter,
whose self-consistent solution in equilibrium plays essential roles
in the large $\mu$ region.
This point would be an advantage to the previous works with NLO
effects~\cite{Faldt1986,BDP,BKR,Bilic:1995tq}.
Thirdly, we discuss the evolution
of the first and second order transition lines and the critical point
with $\beta=2N_c/g^2$.
The finite coupling effects on the critical point have not been
investigated before.
Fourthly, the NLO contribution is expressed
as modifications of the constituent quark mass,
chemical potential and the quark wave function renormalization factor.
Hence the mechanism of the phase diagram evolution becomes clear.

While we are working based on the strong coupling expansion with NLO effects,
we expect the present work would give a valuable picture
in understanding the QCD phase diagram in the real world
through the relation with the MC simulations.
Since the SC-LQCD is based on the same formulation 
as lattice MC simulations,
its results should be consistent with MC results
as long as the applied approximations are valid.
This speculation is supported by previous works on the hadron mass
spectrum~\cite{Kawamoto:1982gw,KlubergStern:1982bs,Jolicoeur:1983tz}.
Very recently, 
the structure of the phase diagram suggested in SC-LQCD
is qualitatively confirmed by a lattice MC simulation
in the strong coupling limit~\cite{deForcrand:2009dh}
based on a Monomer-Dimer-Polymer formulation~\cite{Karsch:1988zx}.
Thus the phase diagram in the strong coupling limit is established
from both side of analytic and numerical studies,
and provide a good starting point to explore the true phase diagram
by evaluating finite coupling effects in the strong coupling expansions.
We find MC studies using one species of unrooted staggered quarks,
and the results around $\beta\sim 5$ have been
extensively discussed~\cite{D'Elia:2002gd,Fodor:2001au}.
In order to compare the SC-LQCD results with those in MC simulations,
we discuss the results in the region $\beta\leq6$
expecting that these $\beta$ values are within the conversion radius.

Although the number of flavors ($N_f=4n_f=4$) 
used in the present work
is different from the real world ($N_f=2+1$),
we could provide valuable results
for the phase diagram investigations.
Flavor dependence of the phase diagrams 
at strong coupling has been studied
by using several species ($n_f=2, 3$)
of staggered fermions~\cite{BKR,Nishida:2003fb},
and we find that the phase diagrams with $n_f=2~\mathrm{and}~3$ are 
qualitatively the same as that with $n_f=1$.
The critical chemical potentials 
at $T=0$ ($\mu_{c,T=0}$) are found to be almost the same.
The critical temperatures at $\mu=0$ ($T_{c,\mu=0}$)
are found to be around 1.2~\cite{BKR,Nishida:2003fb} and 1.06~\cite{BKR}
for $n_f=2$ and 3, respectively.
These values differ from the result of $n_f=1$ ($T_{c,\mu=0}=5/3$)
by 30--40 \%, but the obtained phase diagram structure is very similar.
% BKR92
% n_f  \gamma_c    T_c=\gamma_c^2/N_\tau
% 1    2.579E+00   1.66
% 2    2.190E+00   1.199
% 3    2.060E+00   1.0609
% Nishida
% n_f T_c
% 2   1.22
%
The flavor dependence of the phase diagram is found to be moderate
also in the continuum region.
In Ref.~\cite{D'Elia:2002gd},
the small $\mu$ region of phase diagram 
is investigated by using
MC simulations with four flavor staggered quarks,
and the results are compared with two flavor results~\cite{AC}.
The difference between the phase boundaries
in two and four flavor cases is at most
$7$ \% within a region $N_c\mu<500$ MeV~\cite{D'Elia:2002gd}.
Thus, a ``shape'' of the phase boundary
may not be crucially affected by the flavor effects.
It should be noted that the number of flavors is important
to some of the key features of the phase diagram,
such as the order of the phase transition
and the position of the critical point.

The organization of this paper is as follows.
In Sec.~\ref{sec:Seff}, we provide a brief review 
on the strong coupling ($1/g^2$),
the large dimensional ($1/d$)
and cluster expansions,
and derive the effective action
including the $\mathcal{O}(1/g^2,1/d^0)$ effects.
In Sec.~\ref{sec:Feff},
we derive an analytic expression of the effective potential.
In Sec.~\ref{sec:PD},
we investigate the phase diagram evolution with $\beta$,
and focus on the mechanisms of the
critical temperature and chemical potential modifications.
The ``partially chiral restored (PCR) matter''
is found to appear in the high density region,
and we also discuss its origin.
Finally we summarize our work in Sec.~\ref{sec:CR}.
All through this paper,
we use the lattice unit $a=1$,
and physical values are normalized by $a$.

%%%%%%%%%%%%%%%%%%%%%%%%%%%%%%%%%%%%%%
\section{Effective Action}\label{sec:Seff}

%%%%%%%%%%%%%%%%%%%%%%%%%%%%%%%%%%%%%%
\subsection{Lattice QCD action}\label{subsec:action}
We start from the lattice QCD action and the partition function
with one species of staggered fermion $\chi$
with a quark mass $m_0$.
Gluons are represented 
by the temporal link ($U_0$) 
and spatial link ($U_j,j=1,2,\cdots,d$) variables,
\begin{align}
Z=& \int\mathcal{D}[\chi,\bar{\chi},U_0,U_j]
	\exp\left[ -S_\mathrm{LQCD} \right]
\label{eq:Z}\ ,\\
S_\mathrm{LQCD}
	=& S_F^{(\tau)} +\sum_x m_0 M_x +S_F^{(s)}+S_G
\ ,\\
S_F^{(\tau)}
	=&\frac12\sum_x\left[
		e^\mu\chibar_x U_0 \chi_{x+\hat{0}}
		-e^{-\mu}\chibar_{x+\hat{0}} U^\dagger_0 \chi_x
	\right]
\nn\\
	\equiv&\frac12\sum_x\left[\Vp_x(\mu)-\Vm_x(\mu)\right]
\ ,\\
S_{F}^{(s)}
	=&
		\sum_{x,j}
		\frac{\eta_{j,x}}{2}
		\bigl[
			\bar{\chi}_{x}U_{j,x}\chi_{x+\hat{j}}-(h.c)
		\bigr]
	\equiv \sum_{x,j} s_{j,x}
\ ,\label{eq:SFs}\\
S_G
	=& \frac{2N_c}{g^2} \sum_P
		\Bigl[
1-\frac{1}{2N_c}
\bigl[U_P+U_P^{\dagger}\bigr]
\Bigr]
\ .\label{eq:SG}
\end{align}
Here the trace of the plaquette $U_P$
is defined as,
\begin{align}
U_{P=\mu\nu,x}=\mathrm{tr}_c\left[
	U_{\mu,x}U_{\nu,x+\hat{\mu}}
	U^\dagger_{\mu,x+\hat{\nu}}U^\dagger_{\nu,x}
	\right]
\ .
\end{align}
%%%%%%%%%%%%%%%%%%%%%%%%%%%
In this action, $M_x$ denotes 
the mesonic composite, $M_x=\chibar_x\chi_x$,
and we have defined two other mesonic composites,
$V^\pm$, which contain the temporal link variables.
Sums over color indices are assumed.
Quark chemical potential on the lattice ($\mu$)
is introduced as a weight of the temporal hopping
in the exponential form \cite{Hasenfratz:1983ba},
and the staggered phase factor 
$\eta_{j,x}=(-1)^{x_0+\cdots +x_{j-1}}$
in the spatial action is related to
the Dirac's $\gamma$ matrices 
\cite{Kawamoto:1981hw,KlubergStern:1983dg}.
By using a $\gamma_5$-related 
factor $\epsilon_x=(-1)^{x_0+\cdots+x_{d}}$,
a staggered chiral transformation is given as
$\chi_x\to e^{i\theta\epsilon_x}\chi_x$
\cite{Susskind:1976jm,Sharatchandra:1981si,Kawamoto:1981hw}.
The lattice kinetic action $S_F^{(\tau,s)}$
is invariant under this chiral transformation
in the chiral limit $m_0\to 0$.

Throughout the paper, we consider
the color SU($N_c=3$) case in $3+1$ dimensions ($d=3$).
Temporal and spatial lattice sizes are $N_\tau$ and $L$, respectively.
While $T=1/N_{\tau}$ takes discrete values, 
we consider $T$ as a continuous valued temperature.
%%%T and mu
We take account of finite $T$ effects
by imposing periodic and anti-periodic 
boundary conditions on link variables and quarks, respectively.
We take the static and diagonalized gauge (called Polyakov gauge)
for temporal link variables with respect for the periodicity~\cite{DKS}.

\subsection{Spatial link integral in the strong coupling limit}\label{subsec:SCExp}
In the finite temperature ($T$) treatment,
we obtain the effective action of quarks ($\chi,\bar{\chi}$)
and temporal link variable ($U_0$)
by integrating out the spatial link variables ($U_j$).
We shall evaluate the spatial partition function,
\begin{align}
Z^{(s)}=\int\mathcal{D}U_j
\exp\Bigl[-S_F^{(s)}-S_G\Bigr]\ ,\label{Eq:Zs}
\end{align}
and integrate out the spatial link variables ($U_j$).
In the strong coupling region ($g\gg 1$),
we can treat the plaquette action term 
($S_G\propto 1/g^2$)
through the expansion in the power series of $1/g^2$
(strong coupling expansion).

In the strong coupling limit (SCL), we can omit $S_G$,
and the spatial partition function 
is decomposed into that on each link,
\begin{align}
Z_{\mathrm{SCL}}^{(s)}
	=\int\mathcal{D}U_j~e^{-S_F^{(s)}}
	= \prod_{j,x}\biggl[
		\int dU_j\exp\bigl[-s_{j,x}\bigr]
	\bigg]
\ .\label{Eq:Z_SCL}
\end{align}
We can carry out the $U_j$ integral on each link $(j,x)$
by utilizing the one-link SU($N_c$) group integral formulae~\cite{Gint},
\begin{eqnarray}
&& \int dU~ U_{ab} U^\dagger_{cd} = {1\over N_c} \delta_{ad}\delta_{bc}
\ ,\\
&& \int dU~ U_{ab}U_{cd} \cdots U_{ef}
	= \frac{1}{N_c!} \epsilon_{ac\cdots e} \epsilon_{bd\cdots f}
\ ,\label{Eq:Uint}
\end{eqnarray}
and other higher order integral formulae.
The spatial part of the hadronic effective action density
is obtained as~\cite{Hoek:1981uv},
\begin{align}
&\int dU_j\exp\bigl[-s_{j,x}\bigr]
=\exp\bigl[-s_{j,x}^{(\mathrm{eff})}\bigr]\ ,
\label{Eq:seffjx}\\
&s_{j,x}^{(\mathrm{eff})}
\equiv
\sum_{n=1}^{N_c}
A_n\bigl(
M_xM_{x+\hat{j}}
\bigr)^{n}\nn\\
&\qquad +
A_{j,x}
\bigl(\bar{B}_xB_{x+\hat{j}}
+(-1)^{N_c}(h.c.)\bigr)
\ ,\label{Eq:zjx}
\end{align}
where
$B_x=\epsilon^{ab\cdots c}(\chi^a\chi^b\cdots\chi^c)_x/N_c!$
represent baryonic composites.
Coefficients $(A_n,A_{j,x})$ are summarized in Table~\ref{Tab:Cnjx}
for $N_c=3$.
The spatial partition function in SCL is obtained as,
\begin{align}
Z_{\mathrm{SCL}}^{(s)}
=
\exp\bigl[-\sum_{j,x}s_{j,x}^{(\mathrm{eff})}\bigr]
\ .\label{Eq:ZSCLfin}
\end{align}
%=========================================================================
\begin{table}[tb]
\caption{The coefficients of the hadronic composites
in the effective action at the strong coupling limit.
Detailed explanation to calculate these coefficients are found
in Ref.~\cite{Gint}.
}\label{Tab:Cnjx}
\begin{center}
\begin{tabular}{c|c}
\hline
Coefficients&Values ($N_c=3$)\\
\hline
$A_1$        &$-1/(4N_c)$\\
$A_2$        &$-(N_c^2\cdot (N_c-2)!-N_c!)/(32\cdot N_c^2\cdot N_c!)$\\
$A_3$        &$-(2\cdot N_c!-N_c^3\cdot (N_c-2)!)/(128\cdot N_c^4\cdot N_c!)$\\
$A_{j,x}$    &$-(-1)^{N_c(N_c-1)/2}\eta_{j,x}^{N_c}/2^{N_c}$\\
\hline
\end{tabular}
\end{center}
\end{table}
%=========================================================================

\begin{figure}[bt]
\Psfig{8.5cm}{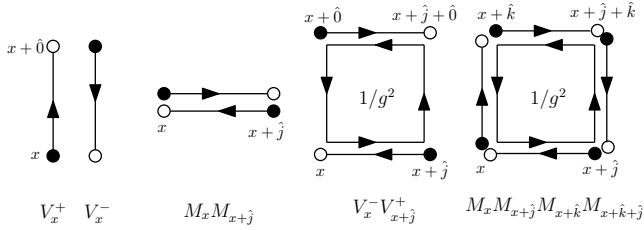}
\caption{Effective action terms
in the strong coupling limit and $1/g^2$ corrections.
Open circles, Filled circles, and arrows show $\chi$, $\chibar$, and
$U_\nu$, respectively.}
\label{Fig:diagrams}
\end{figure}

The sum over spatial directions $\sum_j$ in Eq.~(\ref{Eq:ZSCLfin})
would give rise to a factor $d$
due to the spatial isotropy.
Suppose that the action
$\sum_{j,x}s_{j,x}^{(\mathrm{eff})}$ stays finite at large $d$,
the quark field ($\chi,\bar{\chi}$) should scale as $d^{-1/4}$.
As a result, the mesonic hopping term $\sum_jM_xM_{x+\hat{j}}$ remains
finite $\mathcal{O}(1/d^0)$, 
while higher power terms of quarks are found to be suppressed
as ${\cal O}(1/\sqrt{d})$ for $N_c \geq 3$.
This is called the systematic $1/d$ expansion,
which is proposed first in the application
to the Ising model~\cite{Fisher:1964}.
A spin exchanging term $\sum_j S_xS_{x+\hat{j}}$
is assumed to be finite at large $d$,
and the mesonic hopping $\sum_j M_xM_{x+\hat{j}}$
could be analogue of that \cite{KlubergStern:1982bs}.
In the leading order of the $1/d$ expansion,
the SCL spatial partition function becomes,
\begin{align}
Z_{\mathrm{SCL}}^{(s)}
=
\exp\Bigl[\frac{1}{4N_c}\sum_{j,x}M_xM_{x+\hat{j}}
+\mathcal{O}(1/\sqrt{d})\Bigr]
\ .
\end{align}
In the third diagram of Fig.~\ref{Fig:diagrams},
we display the leading order diagram 
of the $1/d$ expansion.

%=========================================================================
%=========================================================================
\subsection{Strong coupling and cluster expansion}
In order to evaluate the plaquette contribution $S_G$,
it is useful to define an expectation value,
\begin{align}
\big\langle\mathcal{O}\big\rangle
=\frac{1}{Z_{\mathrm{SCL}}^{(s)}}
\int\mathcal{D}U_j~\mathcal{O}[U_j]~e^{-S_{F}^{(s)}}
\ ,
\end{align}
which has a normalization property
$\langle \mathbf{1}\rangle =1$.
The full spatial partition function Eq.~(\ref{Eq:Zs})
can be expressed as,
\begin{align}
Z^{(s)}=Z_{\mathrm{SCL}}^{(s)}
\big\langle e^{-S_G}\big\rangle
\ .\label{Eq:Zs2}
\end{align}
It is well known that the expectation value of the exponential form operator 
with a small factor ({\em i.e.} $1/g^2$) can be evaluated
by using the cumulant expansion~\cite{cumulant},
\begin{align}
\big\langle e^{-S_G} \big\rangle
=&\sum_{n=0}^\infty \frac{(-1)^n}{n!} \big\langle{S_G}^n\big\rangle
\nn\\
=&\exp\biggl[\sum_{n=1}^{\infty}
\frac{(-1)^n}{n!}
\big\langle S_G^n \big\rangle_c
\biggr]
\ ,\label{Eq:cum}
\end{align}
where $\langle\cdots\rangle_c$ is called a cumulant,
and corresponds to the correlation part in the connected diagram contributions,
e.g. $\VEV{S_G^2}_c=\VEV{S_G^2}-\VEV{S_G}^2$.
We find that the effective action from plaquettes is expressed
in terms of cumulants as,
\begin{align}
\Delta S_\mathrm{eff}
	\equiv & - \log\big\langle e^{-S_G} \big\rangle
	= - \sum_{n=1}^\infty \frac{(-1)^n}{n!} 
	\big\langle S_G^n \big\rangle_c
	\ .\label{Eq:Scum}
\end{align}
The $n$-th term in rhs is proportional to $1/g^{2n}$,
and we can identify
$n=1$ term as the next-to-leading order (NLO) effective action,
and $n=2$ term as the next-to-next-to-leading order (NNLO) effective action.

The above identification of the effective action and the strong coupling order
is consistent with the cluster expansion.
In the first line of Eq.~(\ref{Eq:cum}),
average of $S_G^n$ is decomposed into cumulants as,
\begin{align}
&\big\langle S_G^n \big\rangle =
\sum_\mathrm{partition}
\prod_\alpha
\big\langle S_G^{n_\alpha} \big\rangle_c
\nn\\
&=\VEV{S_G}_c^n + \frac{n!}{2!(n-2)!}\VEV{S_G}_c^{n-2}\VEV{S_G^2}_c
	+ \cdots + \VEV{S_G^n}_c \ ,\label{Eq:aveSGn}
\end{align}
where the sum runs over all partitions
satisfying $\sum_\alpha n_\alpha=n$.
The plaquette action $S_G$ is proportional to
a large volume factor $\sum_x\sim\mathrm{Vol.}$
and a small coupling factor $1/g^2$,
hence it is necessary to count both of them.
The first term in Eq.~(\ref{Eq:aveSGn}) is estimated as,
\begin{align}
\big\langle S_G\big\rangle_c^n
\propto \left[\frac{1}{g^2}\sum_x\right]^n
\sim\mathcal{O}(1/g^{2n},\mathrm{Vol.}^n)\ .
\end{align}
In comparison, other terms have smaller powers in volume.
For example, the cumulant of the $n$-th power operator
is proportional to $(\mathrm{Vol.})^1$,
\begin{align}
\big\langle S_G^{~n}\big\rangle_c
\propto
\sum_{\{x_i\}\in \mathrm{conn.}}\biggl[\frac{1}{g^{2n}}\biggr]
\sim\mathcal{O}(1/g^{2n},\mathrm{Vol.}^1)
\ ,
\end{align}
where ``$\{x_i\}\in \mathrm{conn.}$'' represents that
the sum over $\{x_i|i=1,\cdots,n\}$ 
is restricted to connected diagrams,
and such a summation is
$\mathcal{O}(\mathrm{Vol.}^1)$.
In a fixed order of $1/g^{2n}$, 
$\big\langle S_G \big\rangle_c^n$
gives the leading order contribution
in terms of the volume
in the thermodynamical limit, $\mathrm{Vol.} \to \infty$.
Resumming all leading order contributions in volume of the connected diagrams,
we obtain the exponential form shown in the rhs of
Eq.~(\ref{Eq:Scum}).
This resummation corresponds to the so-called cluster expansion~\cite{cumulant},
and is consistent with the strong coupling expansion of the effective action
presented in Eq.~\ref{Eq:Scum}.

\subsection{NLO Effective action}

The NLO contribution
to the effective action Eq.~(\ref{Eq:Scum}) is found to be,
\begin{align}
\Delta S_\mathrm{NLO}
=\big\langle S_G\big\rangle_c
=-\frac{1}{g^2}\sum_P\big\langle U_P+U_P^{\dagger}\big\rangle_c
\ .\label{Eq:cumNLO}
\end{align}
As long as we consider NLO terms,
the cumulant $\langle\cdots\rangle_c$
is equivalent to 
the expectation value $\langle\cdots\rangle$,
\begin{align}
\langle U_P\rangle_c = \langle U_P\rangle 
	= \bigl(1/Z_{\mathrm{SCL}}^{(s)}\bigr)
	\int\mathcal{D}U_j~U_P~e^{-S_{F}^{(s)}}
\ .\label{Eq:cumUp}
\end{align}
We decompose the spatial kinetic action
$S_F^{(s)}$ to the plaquette related and non-related
part,
\begin{align}
S_F^{(s)}
\to&
s_{P}+\sum_{(j,x)\notin P}s_{j,x}\ ,\\
s_{P}
=&
\sum_{(j,x)\in P} s_{j,x}
\ .
\end{align}
In the above sum $\sum_P$,
four links, $(j,x),(k,x+\hat{j}),(j,x+\hat{k}),(k,x)$
are included in a spatial plaquette
$P=(jk,x)$ as shown in the fifth diagram of Fig.~\ref{Fig:diagrams}.
The link integral in Eq.~(\ref{Eq:cumUp})
is also decomposed as follows,
\begin{align}
\langle U_P\rangle_c
=&
\left[ \prod_{(j,x)\in P} e^{s^\mathrm{(eff)}_{j,x}} \right]
\biggl[
\int dU_P~U_P\exp\bigl[-s_{P}\bigr]
\biggr]\nn\\
\times&
\prod_{(k,y)\notin P}
\biggl[
e^{s^\mathrm{(eff)}_{k,y}}
\int dU_{k,y}~\exp\bigl[-s_{k,y}\bigr]
\biggr]
\ .\label{Eq:decompDUj}
\end{align}
The second line shows the plaquette non-related part,
and is found to be unity from the definition of $s^\mathrm{(eff)}_{j,x}$
shown in Eq.~(\ref{Eq:seffjx}).
The prefactor in the first line corresponds to the normalization factor
in the plaquette related part.
This factor is also found to be unity in the leading order of the
$1/d$ expansion, since $s^\mathrm{(eff)}_{j,x}$ contains four quarks,
and is proportional to $d^{-1}$ at large $d$.

The $U_P$ integral part
in Eq.~(\ref{Eq:decompDUj})
contains at most only four links,
and we can perform the link integrals
by using the group integral formulae
Eq.~(\ref{Eq:Uint}) again.
For a temporal plaquette,
$P=(j0,x)$,
we find~\cite{Jolicoeur:1983tz,Faldt1986},
%%%%%%%%%
\begin{align}
\big\langle U_{j0,x}\big\rangle_c
=&
-\frac{1}{4N_c^2} \Vm_x(\mu) \Vp_{x+\hat{j}}(\mu)
+\mathcal{O}\Bigl(d^{-3/2}\Bigr)
\ ,\label{Eq:VV}
\end{align}
where the first term in Eq.~(\ref{Eq:VV})
corresponds to the leading order in the $1/d$ expansion
and shown in the fourth diagram of Fig.~\ref{Fig:diagrams}.
Note that the temporal link variable $U_0$ remains
in the non-local color singlet composites $V^{\pm}$,
and will be integrated out later.
For a spatial plaquette,
$P=(jk,x)$,
$U_P$ integral in Eq.~(\ref{Eq:decompDUj}) is found to be
\cite{Jolicoeur:1983tz,Faldt1986},
\begin{align}
\big\langle U_{jk,x}\big\rangle_c
=\frac{1}{16N_c^4}
M_{x}M_{x+\hat{j}}M_{x+\hat{k}}M_{x+\hat{k}+\hat{j}}
+\mathcal{O}\Bigl(d^{-5/2}\Bigr)\ ,\label{Eq:MMMM}
\end{align}
where $MMMM$ term is the leading order contribution
and is illustrated in the fifth diagram of Fig.~\ref{Fig:diagrams}.

Substituting Eqs.~(\ref{Eq:VV}) and (\ref{Eq:MMMM})
in Eq.~(\ref{Eq:cumNLO}),
we obtain the NLO effective action from plaquettes,
$\Delta S_\mathrm{NLO}$.
The plaquette sum $\sum_P$ in $\Delta S_\mathrm{NLO}$ leads to
$\sum_j\sim d$ and $\sum_{jk}\sim d^2$
for temporal and spatial plaquettes, respectively.
Since the quark fields $(\chi,\bar{\chi})$
scales as $d^{-1/4}$,
the composites in Eq.~(\ref{Eq:VV}) and (\ref{Eq:MMMM})
scale as $\Vm\Vp \sim d^{-1}$ and $MMMM\sim d^{-2}$.
Putting all together,
$\Vm\Vp$ and $MMMM$ give $\mathcal{O}(1/d^0)$ contributions
in $\Delta S_\mathrm{NLO}$.

In the following, we consider
the leading (SCL) and the NLO
in the strong coupling expansion,
and the leading order of the $1/d$ expansion.
The effective action is found to be,
\begin{align}
S_{\mathrm{eff}}
	=& S_\mathrm{SCL} 
	+ \Delta S^{(\tau)}
	+ \Delta S^{(s)}
	+ \mathcal{O}(1/\sqrt{d}, 1/g^4)
\label{eq:Seff}\ ,\\
S_\mathrm{SCL} 
        =&\frac12\sum_x\left[\Vp_x(\mu)-\Vm_x(\mu)\right]
        + m_0\sum_x M_x
\nn\\
        &- \frac{1}{4N_c}\sum_{x,j>0} M_xM_{x+\hat{j}}\ ,\\
\Delta S^{(\tau)}
	=&\frac{\beta_{\tau}}{4d}
		\sum_{x,j>0} 
		\left[
		 \Vp_x(\mu) \Vm_{x+\hat{j}}(\mu)
		+\Vp_x(\mu) \Vm_{x-\hat{j}}(\mu)
		\right]
\label{Eq:ActionGT}\ ,\\
\Delta S^{(s)}
	=&\frac{-\beta_s}{d(d-1)}
		\sum_{x,0<k<j} 
		M_{x}
		M_{x+\hat{j}}
		M_{x+\hat{k}}
		M_{x+\hat{k}+\hat{j}}
	\ ,
\label{Eq:ActionGS}
\\
\beta_{\tau}=&\frac{d}{N_c^2g^2}\ ,\quad
\beta_s=\frac{d(d-1)}{8N_c^4g^2}\ ,
\end{align}
where
$\Delta S^{(\tau)}$ and $\Delta S^{(s)}$
come from the temporal and spatial plaquettes
shown in Eq.~(\ref{Eq:VV}) and (\ref{Eq:MMMM})
including their hermit conjugates,
respectively.
The considered contributions 
$S_F^{(\tau)}$, $MM$, and $\Delta S^{(\tau,s)}$
are summarized in Fig.~\ref{Fig:diagrams}.

%----------------------------------------------------------------------*
\section{Effective Potential}\label{sec:Feff}

The effective action derived in the previous section
still contains quark fields $(\chi,\bar{\chi})$
and temporal link variables $U_0$.
In this section, we obtain the effective potential $\Feff{eff}$
by integrating out these variables
in the mean field approximation,
\begin{eqnarray}
\int\mathcal{D}[U_0,\chi,\bar{\chi}]~e^{-S_\mathrm{eff}}
=&\int\mathcal{D}[\Phi]~e^{-N_{\tau}L^d\Feff{eff}[\Phi]}
\nn\\
\approx& e^{-N_{\tau}L^d\Feff{eff}[\Phi]}\bigl|_\mathrm{stationary}
\ .
\end{eqnarray}
In this step, several auxiliary fields ($\Phi$) 
including the chiral condensate
$\sigma$ are introduced on the same footing,
and the NLO effects are self-consistently evaluated.

\subsection{Effective potential in the strong coupling limit}
\label{subsec:SCLim}

Before discussing the NLO effects,
we briefly summarize the procedure 
to obtain the effective potential in SCL.
The effective action $S_\mathrm{SCL}$ contains
the chirally invariant four-fermi term $MM$.
We apply the so-called Hubbard-Stratonovich (HS) transformation.
The four fermi term $MM$ is reduced to bilinear forms
in $\chi$ and $\chibar$
by performing the Gaussian transformation with a auxiliary field $\sigma$,
\begin{align}
&\exp\Biggl[~\sum_{x,j>0} \frac{M_xM_{x+\hat{j}}}{4N_c}\Biggl]
=e^{b_\sigma\sum_{xy} M_xV_{xy}M_{y}/2}
\nn\\
&= \int {\cal D}\sigma~
 e^{-(b_\sigma/2)\sum_{xy}\bigl[(\sigma+M)_xV_{xy}(\sigma+M)_y-M_xV_{xy}M_{y}\bigr]}\nn\\
&= \int {\cal D}\sigma\,
	\exp\Biggl[
	-\sum_{xy}b_\sigma\left(
			\frac{\sigma_xV_{xy}\sigma_y}{2}+\sigma_x V_{xy}M_y
		\right)
	\Biggr]
\nn\\
&\approx
\exp\biggl[-N_\tau L^d\frac{b_\sigma}{2}\sigma^2
-b_\sigma\sigma \sum_x M_x\biggr]\equiv e^{-S_{\sigma}}\ ,
\label{Eq:sigma}
\end{align}
where $b_\sigma=d/2N_c$,
and the matrix
$V_{xy}=\sum_j(\delta_{x+\hat{j},y}+\delta_{x-\hat{j},y})/2d$
represents the meson hopping.
In the last line,
$\sigma$ is assumed to be a constant,
which is determined by the stationary condition
$\partial \Feff{eff}^{\mathrm{SCL}}/\partial \sigma=0$.
Under this condition, 
the auxiliary field $\sigma$ is found to be
the chiral condensate $\sigma=-\sum_x\langle M_x\rangle/(N_{\tau}L^d)$.
Thus the non-linear term $MM$ is converted 
to the quark mass term $b_\sigma\sigma\chibar\chi$,
where the finite chiral condensate $\sigma$
spontaneously breaks the chiral symmetry
and generates the quark mass dynamically.

Now the total effective action
reduces to a bilinear form of $(\chi,\chibar)$,
\begin{align}
S_\mathrm{SCL}
\simeq&
\sum_x\left[\frac{\Vp_x(\mu)-\Vm_x(\mu)}{2}+m_qM_x\right]
+N_\tau L^d \frac{b_\sigma}{2}\sigma^2
\nn\\
\simeq&
\sum_{xy}\bar{\chi}_xG_{xy}^{-1}(m_q,\mu)\chi_y
+N_{\tau}L^d 
\frac{b_{\sigma}}{2}\sigma^2\ ,
\label{eq:SSCL_fin}
\end{align}
where $m_{q}=m_0+b_{\sigma}\sigma$
represents the constituent quark mass,
and the inverse propagator of quarks is given as,
\begin{align}
&G_{xy}^{-1}(m_q,\mu)\nn\\
&=
\frac{\delta_{\mathbf{xy}}}{2}
\Bigl(
e^{\mu}U_{0,x}\delta_{x+\hat{0},y}
-e^{-\mu}U_{0,x}^{\dagger}\delta_{x-\hat{0},y}
\Bigr)
+ m_{q}\delta_{xy}
\label{eq:qhop}\ ,
\end{align}

We take account of finite $T$ effects
by imposing periodic and anti-periodic 
boundary conditions on link variables and quarks, respectively.
We take the static and diagonalized gauge (called the Polyakov gauge)
for temporal link variables with respect for the periodicity~\cite{DKS},
\begin{eqnarray}
U_{0}(\tau,\mathbf{x})
=
\mathrm{diag}
(e^{i\theta_1(\mathbf{x})/N_\tau},\cdots,e^{i\theta_{N_c}(\mathbf{x})/N_\tau})
\ .\label{eq:Tgauge}
\end{eqnarray}
The corresponding Haar measure is given in the form of the Van der Monde
determinant,
\begin{align}
\int dU_0
\equiv&
\left[
\prod_{a=1}^{N_c}
\int_{-\pi}^{\pi}
\frac{d\theta_a}{2\pi}
\right]
\prod_{a<b}
\bigl|
e^{i\theta_a}-e^{i\theta_b}
\bigr|^2\nn\\
&\times 
2\pi\delta\Bigl(\sum_a\theta_a\Bigr)
\ ,\label{eq:PHaar}
\end{align}
where the delta function reflects the $SU(N_c)$ property,
{\em i.e.} the baryonic effect in the temporal direction.
Owing to the static property of the auxiliary field $\sigma$
and the temporal link variable
in the Polyakov gauge,
the partition function
$Z_q=\int_{\chi,\bar{\chi},U_0}~e^{-\bar{\chi}G^{-1}\chi}$
is completely factorized in terms of
the frequency modes. Hence the quark path integral
can be done in each mode independently,
and leads to the simple product in the frequency ($\prod_{\omega}$).
By utilizing the Matsubara method
(see for example the appendices 
in Refs.~\cite{Nishida:2003uj,Kawamoto:2005mq}),
we obtain the partition function as,
\begin{align}
Z_q=&
\prod_{\mathbf{x}}
\Biggl[
\int dU_{0\mathbf{x}}~
e^{E_q/T}
\prod_{a=1}^{N_c}
\Bigl[
1+e^{-(E_q-\mu)/T+i\theta_{\mathbf{x}}^a}
\Bigr]\nn\\
&\times\Bigl[
1+e^{-(E_q+\mu)/T+i\theta_{\mathbf{x}}^a}
\Bigr]
\Biggr]\ ,\label{Eq:ZqU0}
\end{align}
where $N_{\tau}^{-1}$ is identified as temperature $T$,
and
$E(m_{q}(\sigma))=\sinh^{-1}\bigl[m_{q}(\sigma)\bigr]$
corresponds to the quark excitation energy.
Substituting Eq.~(\ref{eq:PHaar}) for Eq.~(\ref{Eq:ZqU0}),
the remnant $U_0$ integral can be carried out
in a straightforward manner~\cite{DKS}
(the explicit procedure is summarized 
in the appendix in Ref.~\cite{Nishida:2003fb}).
The resultant effective potential \cite{DKS,Faldt1986}
is a function of the chiral condensate $\sigma$,
temperature $T$ and quark chemical potential $\mu$,
\begin{align}
\Feff{eff}^{\mathrm{SCL}}(\sigma;T,\mu)
=&\frac{b_{\sigma}}{2}\sigma^2+\Vq(m_{q}(\sigma);T,\mu)
\label{eq:Feff_scl}\ ,\\
\Vq(m_{q};T,\mu)
=&-T\log\left[
	X_{N_c}(m_q)+2\cosh\left[\frac{N_c\mu}{T}\right]
	\right]
\ ,\label{eq:Vq_scl}\\
X_{N_c}(m_q)=&
\frac{\sinh\bigl[(N_c+1)E_q(m_q)/T\bigr]}{\sinh\bigl[E_q(m_q)/T\bigr]}
\ ,
\end{align}
The same result is also obtained by another
method~\cite{Faldt1986},
by utilizing recursion formulae.
The phase diagram is obtained by performing the
minimum search of the effective potential $\Feff{eff}^{\mathrm{SCL}}$,
and its structure has been investigated
in Refs.~\cite{BDP,BKR,Bilic:1995tq,Fukushima:2003vi,Nishida:2003fb,Kawamoto:2005mq}.

\subsection{Extended Hubbard-Stratonovich Transformation}\label{subsec:EHS}

We shall now evaluate NLO correction terms $\Delta S^{(\tau,s)}$
in the effective action Eq.~(\ref{eq:Seff}) in the mean field approximation.
The temporal plaquette term $\Delta S^{(\tau)}$ is composed of
the product of different composites
$\Vp_x$ and $\Vm_{x\pm\hat{j}}$.
The standard HS transformation shown in
Eq.~(\ref{Eq:sigma}) cannot be applied
for such a term.
Hence we apply here a recently developed method named
{\em extended Hubbard-Stratonovich (EHS) transformation}~\cite{QY,Lat08_Miura}.
Let us consider to evaluate a quantity 
$e^{\alpha A B}$,
where $(A,B)$ and $\alpha$
represent arbitrary composite fields and a positive constant,
respectively.
We can represent $e^{\alpha AB}$ in the form of Gaussian integral
over two auxiliary fields $(\varphi,\phi)$,
\begin{align}
e^{\alpha A B}
&= \int\, d\varphi\, d\phi\,
	e^{-\alpha\left\{
		 (\varphi-(A+B)/2)^2
		+(\phi-i(A-B)/2)^2
		\right\}+\alpha A B}
\nonumber\\
&= \int\, d\varphi\, d\phi\,
	e^{-\alpha\left\{
		\varphi^2-(A+B)\varphi
		+ \phi^2 - i(A-B)\phi
		\right\}}
\label{Eq:EHSa}
\ .
\end{align}
The integral over the new fields $(\varphi,\phi)$
is approximated by the saddle point value,
$\varphi = \VEV{A+B}/2$ and $\phi = i\VEV{A-B}/2$.
Specifically 
in the case where both $\VEV{A}$ and $\VEV{B}$ are real,
which applies to the later discussion,
the stationary value of $\phi$ becomes pure imaginary.
Thus we replace $\phi \to i\omega$
and require the stationary condition for the real value of $\omega$,
\begin{eqnarray}
e^{\alpha A B}
&\approx& e^{-\alpha\left\{
		\varphi^2-(A+B)\varphi-\omega^2+(A-B)\omega
	\right\}}\Big|_{\mathrm{stationary}}
\label{Eq:EHSc}
\ .
\end{eqnarray}
In the case of $A=B$, Eq.~(\ref{Eq:EHSc}) reduces to
the standard HS transformation.
We find that $e^{\alpha AB}$ is invariant
under the scale transformation, 
$A\to\lambda A$ and $B\to \lambda^{-1}B$.
In our previous work \cite{Kawamoto:2005mq},
a similar invariance exists but
is broken after the saddle point approximation.
As a result, a careful treatment is necessary
in order to determine the explicit value of the parameter.
In the present derivation,
the scale invariance is kept 
in rhs of Eq.~(\ref{Eq:EHSc}),
since the combinations
$\varphi-\omega=\langle A\rangle$
and
$\varphi+\omega=\langle B\rangle$
transform in the same way
as $A$ and $B$, respectively.
This means that the effective potential is independent
of the choice of $\lambda$.

Now we apply EHS to NLO terms.
For the spatial plaquette action terms, $\Delta S^{(s)}$,
we substitute 
$(\beta_s/d(d-1), M_xM_{x+\hat{j}},M_{x+\hat{k}}M_{x+\hat{k}+\hat{j}})$ 
in Eq.~(\ref{Eq:EHSc}), and obtain,
\begin{align}
&\Delta S^{(s)}
\approx\frac{\beta_s}{d(d-1)}
\sum_{x,0<k<j}
\bigl[
\varphi_s^2-\omega_s^2
\nn\\
&
-(\varphi_s-\omega_s)M_xM_{x+\hat{j}}
-(\varphi_s+\omega_s)M_{x+\hat{k}}M_{x+\hat{k}+\hat{j}}
\bigr]
\nn\\
&\approx N_{\tau}L^d\frac12\beta_s\varphi_s^2
-\frac{\beta_s\varphi_s}{d} \sum_{x,j>0} M_xM_{x+\hat{j}}
\label{Eq:EHSsA}
\ .
\end{align}
In the last line, we have assumed that the auxiliary fields
take constant and isotropic values.
Under this constant auxiliary field assumption,
$\omega_s$ effects disappear
and the sum $\sum_{0<k<j}$ leads to a factor $d(d-1)/2$
for the $\varphi_s^2$ term.
As shown in the last line,
the coupling terms of $\varphi_s$ and $M$ are rearranged to
the same form as the meson hopping term in the SCL effective action
by using the translational invariance.
This $M_xM_{x+\hat{j}}\varphi_s$ term
can be absorbed into the meson hopping effects 
$M_xM_{x+\hat{j}}$
in the SCL,
\begin{align}
\frac{b_\sigma}{2d}\sum_{x}M_xM_{x+\hat{j}}
\to
\frac{b_\sigma+2\beta_s\varphi_s}{2d}\sum_{x}M_xM_{x+\hat{j}}
\ .
\end{align}
Thus the spatial NLO contributions lead to a shift of
the coefficient for the meson hopping effects,
which can be evaluated by 
introducing the chiral condensate $\sigma$ via
Eq.~(\ref{Eq:sigma}).
The coefficient modification is cared by
replacing $S_{\sigma}$ in Eq.~(\ref{Eq:sigma}) with,
\begin{align}
\tilde{S}_{\sigma}
=N_\tau L^d\frac{\tilde{b}_\sigma}{2}\sigma^2
+\tilde{b}_\sigma\sigma \sum_x M_x
\ ,
\end{align}
where $\tilde{b}_{\sigma}=d/(2N_c)+2\beta_s\varphi_s$.
The constituent quark mass 
is found to be modified as,
\begin{align}
m^{\prime}_{\mathrm{q}}=m_0+\tilde{b}_{\sigma}\sigma\ .
\end{align}

For the temporal plaquette action $\Delta S^{(\tau)}$,
we substitute
$(\alpha,A,B)=(\beta_{\tau}/4d,-\Vp_x(\mu),\Vm_{x+\hat{j}}(\mu))$, 
and obtain,
\begin{align}
\Delta S^{(\tau)}
\approx& 
\frac{\beta_{\tau}}{4d}\sum_{x,j>0}\Bigl[
		\varphi_\tau^2+\left[\Vp_x(\mu)-\Vm_{x+\hat{j}}(\mu)\right]
		\varphi_\tau
\nn\\
&-
\omega_\tau^2-\left[\Vp_x(\mu)+\Vm_{x+\hat{j}}(\mu)\right]\omega_\tau
	+(j\leftrightarrow -j)\Bigr]
\nn\\
\approx&
N_\tau L^d\frac{\beta_\tau}{2}(\varphi_\tau^2-\wt^2)
\nn\\
+&\frac{\beta_\tau}{2}\sum_x\left[
	(\varphi_\tau-\wt)\Vp_x(\mu)
	-(\varphi_\tau+\wt)\Vm_x(\mu)
	\right]
\ .\label{eq:HSforPt}
\end{align}
In the last line, we again assume that the auxiliary fields 
$\varphi_{\tau}$ and $\omega_{\tau}$ are constant and isotropic,
then $\sum_j$ leads to a factor $d$.
We combine $V^{\pm}$ terms in Eq.~(\ref{eq:HSforPt})
with those in the SCL temporal action $S_F^{(\tau)}$,
and the coefficients of $V^{\pm}$ are found to become,
$Z_\mp/2$, where $Z_\pm =1+\beta_{\tau}(\varphi_{\tau}\pm\omega_{\tau})$.
We rewrite these coefficients as
$Z_\pm = Z_\chi \exp(\pm\delta\mu)$.
Thus the SCL terms $S_F^{(\tau)}$
is modified by the temporal NLO effects as,
\begin{align}
\tilde{S}_{F}^{(\tau)}
	&= \frac12\sum_x
		\left[
			Z_-\Vp_x(\mu)
			-Z_+\Vp_x(\mu)
		\right]
\nn\\
	&= \frac{Z_{\chi}}{2}\sum_x
		\Bigl(
		e^{\mu}e^{-\delta\mu}\bar{\chi}_{x}U_{0,x}\chi_{x+\hat{0}}
		-e^{-\mu}e^{\delta\mu}(h.c.)
		\Bigr)
\nn\\
	&= \frac{Z_\chi}{2}\sum_x
		\left[
			\Vp_x(\tilde{\mu})
			-\Vm_x(\tilde{\mu})
		\right]
\ ,
\end{align}
where,
\begin{align}
Z_{\chi}&=\sqrt{Z_{+}Z_{-}}\ ,\quad
e^{\tilde{\mu}}=e^{\mu}e^{-\delta\mu}=e^{\mu}\sqrt{\frac{Z_-}{Z_+}}\ .
\label{Eq:tildemu}
\end{align}
In this way, temporal NLO contributions are expressed as
the quark wave function renormalization factor $Z_{\chi}$
and the dynamical shift of chemical potential $\delta\mu$.
In Table~\ref{Tab:aux}, we summarize introduced auxiliary fields.
%=========================================================================
\begin{table}[t]
\caption{The auxiliary fields and their stationary values.
In the stationary value of $\varphi_{\tau}$,
$\varphi_0=N_c-Z_{\chi}\tilde{m}_q+\beta_{\tau}\omega_{\tau}^2$.
}\label{Tab:aux}
\begin{center}
\begin{tabular}{c|c|c}
\hline
Aux. Fields&Mean Fields &Stationary Values\\
\hline
$\sigma$        &$\langle -M\rangle$
		&$-(1/Z_{\chi})(\partial\Vq/\partial \tilde{m}_{q})$\\
$\varphi_s$     &$\langle MM\rangle$
		&$\sigma^2$\\
$\varphi_\tau$	&$-\langle (\Vp-\Vm)/2\rangle$
		&$2\varphi_0/(1+\sqrt{1+4\beta_\tau\varphi_0})$\\
$\wt$		&$-\langle (\Vp+\Vm)/2\rangle$
		&$-\partial\Vq/\partial\tilde{\mu}=\rho_q$\\
\hline
\end{tabular}
\end{center}
\end{table}
%=========================================================================
\subsection{Effective potential}\label{subsec:Feff}
Now the effective action
reduces to a bilinear form 
in terms of the quark fields ($\chi,\bar{\chi}$),
\begin{align}
S_\mathrm{eff}
=&
Z_{\chi}
\sum_{xy}\bar{\chi}_x G_{xy}^{-1}(\tilde{m}_q,\tilde{\mu})\chi_y\nn\\
&+
N_{\tau}L^d
\biggl[
\frac{\tilde{b}_{\sigma}}{2}\sigma^2
+\frac{\beta_s}{2}\varphi_s^2
+\frac{\beta_{\tau}}{2}
\bigl(
\varphi_{\tau}^2-\omega_{\tau}^2
\bigr)
\biggr]\ ,
\label{eq:Seff_fin}
\end{align}
where $G_{xy}^{-1}(\tilde{m}_q,\tilde{\mu})$
is given in Eq.~(\ref{eq:qhop}) with
modifications $(m_q,\mu)\to(\tilde{m}_q,\tilde{\mu})$.
We note that the constituent quark mass
is modified again 
due to the quark wave function renormalization factor $Z_{\chi}$,
\begin{align}
\tilde{m}_{q}
=\frac{m_{q}^{\prime}}{Z_{\chi}}
=\frac{m_0+\tilde{b}_{\sigma}\sigma}{Z_{\chi}}
\ .\label{Eq:mq}
\end{align}
The remnant integrals 
$\int_{\chi,\bar{\chi},U_0}~e^{-Z_{\chi}\sum\bar{\chi}G(\tilde{m}_q,\tilde{\mu})\chi}$
can be evaluated in the same manner as in SCL.
The effective potential (free energy density)
is obtained as a function of the auxiliary fields
$\Phi=(\sigma,\varphi_{\tau,s},\omega_{\tau})$,
the temperature $T$ and 
the quark chemical potential $\mu$,
\begin{align}
&\Feff{eff}(\Phi;T,\mu)
=\Feff{aux}(\Phi)
+\Vq(\tilde{m}_{q}(\Phi);T,\tilde{\mu})\ ,
\label{Eq:Feff}\\
&\Feff{aux}(\Phi)
=\frac{\tilde{b}_{\sigma}\sigma^2}{2}
+\frac{\beta_s\varphi_s^2}{2}
+\frac{\beta_{\tau}}{2}
\bigl(
\varphi_{\tau}^2-\omega_{\tau}^2
\bigr)-N_c\log Z_{\chi}\ ,
\end{align}
where $\Vq(\tilde{m}_{q}(\Phi);T,\tilde{\mu})$
has the same functional form as that in SCL Eq.~(\ref{eq:Vq_scl})
except for modifications $m_q\to\tilde{m}_q(\Phi)$
and $\mu\to\tilde{\mu}$.
The additional term $-N_c\log Z_\chi$,
which has no counterpart in the SCL, 
appears from the quark wave function renormalization factor $Z_{\chi}$
through the fermion determinant contribution,
$-\log[\mathrm{det}(Z_\chi G^{-1})]$.

We have introduced four kinds of auxiliary fields
$\Phi=(\sigma,\varphi_s,\varphi_{\tau},\wt)$,
and it may contain
some redundant degrees of freedom.
This can be cared by considering
stationary conditions shown in Eq.~(\ref{Eq:EHSc}),
\begin{align}
\frac{\partial\Feff{eff}}{\partial\Phi}
=
\frac{\partial\Feff{aux}}{\partial\Phi}
+\frac{\partial\Vq}{\partial\tilde{m}_q}
\frac{\partial\tilde{m}_q}{\partial\Phi}
+\frac{\partial\Vq}{\partial\tilde{\mu}}
\frac{\partial\tilde{\mu}}{\partial\Phi}=0
\label{Eq:Stationary}\ .
\end{align}
Note that $\Vq$ depends on the auxiliary fields
via the two dynamical variables $\tilde{m}_q$ and $\tilde{\mu}$.
Substituting $\sigma$ for $\Phi$ in Eq.~(\ref{Eq:Stationary}),
we obtain the relation,
\begin{align}
\sigma = -\frac{1}{Z_{\chi}}\frac{\partial\Vq}{\partial \tilde{m}_{q}}
\ .
\end{align}
By utilizing this result,
the stationary condition for $\varphi_s$
leads to $\varphi_s=\sigma^2$.
Substituting $\varphi_{\tau}$ and $\omega_{\tau}$
for $\Phi$, we obtain a coupled equation
for $\varphi_{\tau}$ and $\omega_{\tau}$,
whose solution is found to be,
\begin{align}
\varphi_\tau
&=
\frac{2\varphi_0}{1+\sqrt{1+4\beta_\tau\varphi_0}}\ ,\\
\varphi_0&=N_c-Z_{\chi}\tilde{m}_q+\beta_{\tau}\omega_{\tau}^2\ ,\\
\omega_\tau
&=
-\frac{\partial\Vq}{\partial\tilde{\mu}}
=-\frac{\partial\Feff{eff}}{\partial\mu}
\label{Eq:rhoq}
\ .
\end{align}
Equation~(\ref{Eq:rhoq}) indicates
the stationary value of $\omega_{\tau}$ is 
nothing but the quark number density $\rho_q$.
The stationary conditions are summarized in Table~\ref{Tab:aux}.

The auxiliary fields $\varphi_{\tau,s}$ 
are found to be explicit functions of $\sigma$ and $\wt$
via stationary conditions,
while $\omega_{\tau}$ becomes a $(T,\mu)$ dependent implicit function,
$\omega_{\tau}=\rho_q(\sigma,\omega_\tau;\mu,T)$.
Hence we need a self-consistent treatment 
in the minimum search of $\Feff{eff}(\sigma,\wt)$
in order to determine vacua.
This is a consequence of
the multi-order parameter ($\sigma,\wt$) treatment,
and a new feature compared
with the previous works~\cite{Faldt1986,BDP,BKR,Bilic:1995tq}.
%=========================================================================

The auxiliary field $\wt$ may be interpreted
as a repulsive vector field for quarks.
In relativistic mean field (RMF) models of nuclei~\cite{RMF},
the isoscalar-vector field $\omega$ contributes
to the energy density as,
\begin{align}
\varepsilon_V=-m_\omega^2 \omega^2/2
+ g_{\omega N}\rhoB(\tilde{\mu}_{\scriptscriptstyle B})\omega
+ \ldots\ ,
\end{align}
where $\omega$ is the temporal component of the omega meson field $\omega^\nu$.
The negative coefficient of $\omega^2$ results in 
the repulsive potential for nucleons, $+g_{\omega N}\omega$,
and the coupling with the baryon density $\rhoB$
leads to the shift of $\mu_\ssB$ as,
\begin{align}
E+g_{\omega}\omega-\mu_\ssB = E-(\mu_\ssB-g_{\omega}\omega)
=E-\tilde{\mu}_\ssB\ .
\end{align}
The saddle point constraint gives $\omega\propto\rhoB$.
Most of these characters apply to the auxiliary field $\wt$
introduced in the present work.
For example, the $\wt$ contribution to the effective action
in Eq.~(\ref{eq:HSforPt}) is rewritten as
$-\beta_\tau\wt^2/2+\beta_\tau\wt\rho_q$,
and the stationary condition is $\wt=\rho_q$.
When we replace quarks with baryons and introduce an appropriate 
scaling factor for $\wt$,
the above two points are consistent with the properties
of $\omega$ vector field in RMF.

\section{Phase diagram evolution}\label{sec:PD}
In the previous section,
we have derived an analytic expression
of the effective potential 
$\Feff{eff}$, which contains effects of the next-to-leading
order (NLO) of the $1/g^2$ expansion.
In this section,
we investigate the phase diagram evolution
with the finite coupling effects $\beta=2N_c/g^2$
based on the effective potential $\Feff{eff}$.
By developing
a self-consistent treatment of two order parameters,
($\sigma,\wt$),
we study the $\beta$ dependence of
the critical temperature, critical chemical potential, the critical point
and the phase diagram.
We also discuss partially chiral restored (PCR) matter.

%---------------------------------------------------------------------
\subsection{Self-consistent treatment in vacuum search}\label{subsec:SelfCons}
%---------------------------------------------------------------------
%---------------------------------------------------------------------
\begin{figure}[tbh]
\Psfig{6cm}{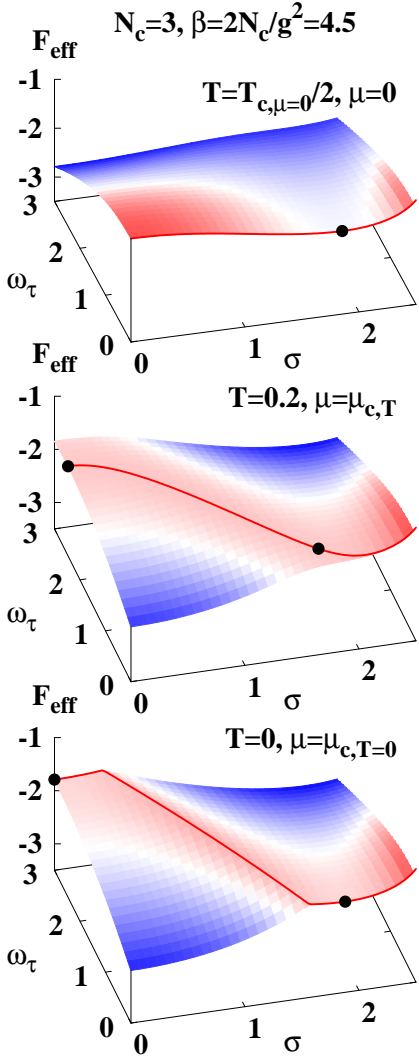}
\caption{(Color online)
The effective potential
as a function of $\sigma$ and $\omega_{\tau}$ at 
$(T,\mu)=(T_{c,\mu=0}/2, 0), (0.2, \mu_{c,T=0.2}), (0, \mu_{c,T=0})$
in the lattice unit.
The solid line represent the set of points
which satisfy the stationary condition Eq.~(\protect\ref{Eq:rhoq}).}
\label{Fig:sigwt}
\end{figure}
%---------------------------------------------------------------------
%---------------------------------------------------------------------
\begin{figure}[tb]
\Psfig{8.0cm}{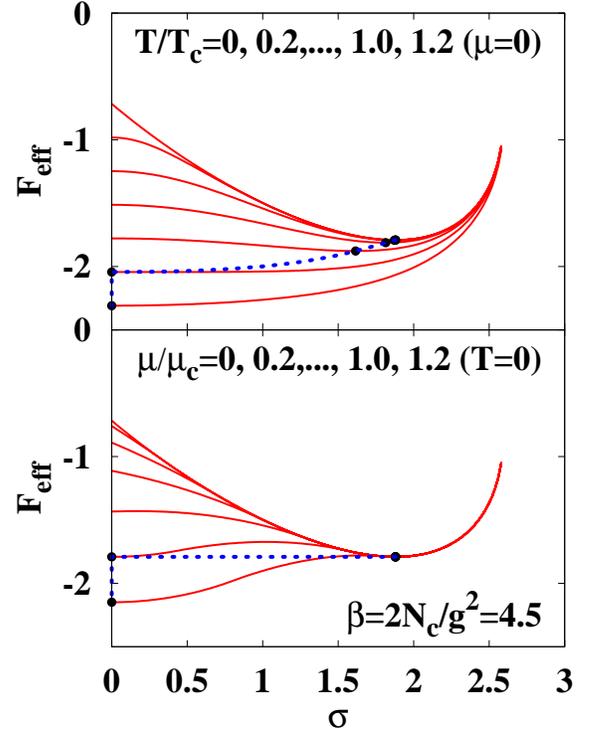}
\caption{%
The effective potential as a function of $\sigma$
on the $T$-axis (upper panel, $\mu=0$)
and on the $\mu$-axis (lower panel, $T=0$)
with $\beta=2N_c/g^2=4.5$
in the lattice unit.
The filled circles represent the minimum points.
}
\label{Fig:surf}
\end{figure}
%---------------------------------------------------------------------

Solving the stationary condition
of $\Feff{eff}$ with respect to $\sigma$ and $\wt$,
\begin{align}
\frac{\partial\Feff{eff}}{\partial\sigma}=
\frac{\partial\Feff{eff}}{\partial\wt}=0
\ ,
\label{Eq:Stat}
\end{align}
corresponds to searching for a saddle point of $\Feff{eff}$
in the $(\sigma,\wt)$ plane.
Since the quark number density $\rho_q$ is an increasing function of
$\tilde{\mu}$ which is a decreasing function of $\wt$,
the stationary condition for $\wt$, $\wt=\rho_q$, has a single solution,
$\wt=\wt^\mathrm{stat.}(\sigma,T,\mu)$,
for a given value of $\sigma$ at finite $T$.
The coefficient of $\wt^2$ is negative in $\Feff{eff}$,
hence
the solution gives a maximum of $\Feff{eff}$ for a given $\sigma$.
Thus the stationary point in $(\sigma,\wt)$ is the saddle point
of $\Feff{eff}$, at which $\Feff{eff}$ is convex downward and upward
in $\sigma$ and $\wt$ directions
($\partial^2\Feff{eff}/\partial\sigma^2>0$
 and $\partial^2\Feff{eff}/\partial\wt^2<0$),
respectively.
Generally, we may have several solutions of Eq.~(\ref{Eq:Stat}),
among which the lowest $\Feff{eff}$ dominates the partition function.

In Fig.~\ref{Fig:sigwt}, 
we show $\Feff{eff}$ as a function of $(\sigma,\wt)$.
Solid lines show the solution of the stationary condition for $\wt$,
$\wt=\wt^{\mathrm{stat.}}$,
and filled circles show the saddle points.
At $\mu=0$,
$\Feff{eff}$ becomes an even function of $\wt$,
and the stationary value of $\wt$ is always zero
as shown in the upper panel of Fig.~\ref{Fig:sigwt}.
At finite $\mu$,
we have to solve the coupled equations (\ref{Eq:Stat}) self-consistently.
The solution $\wt^{\mathrm{stat.}}$ is a smooth function of $\sigma$ at finite $T$
as shown in the middle panel of Fig.~\ref{Fig:sigwt}.
In the case of $T=0$ and finite $\mu$,
$\partial\Feff{eff}/\partial\wt$ is discontinuous at $\tilde{\mu}=E_q$.
This discontinuity comes from the functional form of 
the quark free energy at $T=0$,
\begin{eqnarray}
\Vq(\tilde{m}_q,\tilde{\mu},T=0)=
\begin{cases}
-N_cE_q&(E_q\geq\tilde{\mu})\\
-N_c\tilde{\mu}&(E_q\leq \tilde{\mu})
\end{cases}
\ .\label{eq:Vq_T0}
\end{eqnarray}
The ridge found in the lower panel of Fig.~\ref{Fig:sigwt}
corresponds to the line $\tilde{\mu}=E_q$,
where $\Feff{eff}$ takes a maximum value for a given $\sigma$.
The stationary value $\wt^\mathrm{stat.}$ at finite $T$
approaches this ridge in the limit $T\to 0$.
Thus the stationary condition for $\wt$ is found to be equivalent
to searching for $\wt$ which maximizes $\Feff{eff}$ for each $\sigma$
also at $T=0$.

The effective potential as a function of $\sigma$ for given $(T,\mu)$
is defined as,
$\Feff{eff}(\sigma)=\Feff{eff}(\sigma,\omega=\omega^\mathrm{stat.}(\sigma))$,
whose minimum point corresponds to the equilibrium.
In Fig.~\ref{Fig:surf},
we show $\Feff{eff}(\sigma)$
on the $T$-axis ($\mu=0$)
and on the $\mu$-axis ($T=0$) at $\beta=4.5$, as an example.
The chiral phase transitions in these cases are found to be
the second and first order, respectively, as in the case of SCL.
In the following subsections, we discuss the nature of these phase transitions.

%---------------------------------------------------------------------
\subsection{Critical temperature at zero chemical potential}\label{subsec:Tc}
%---------------------------------------------------------------------

%---------------------------------------------------------------------
\begin{figure}[tbh]
\Psfig{8.0cm}{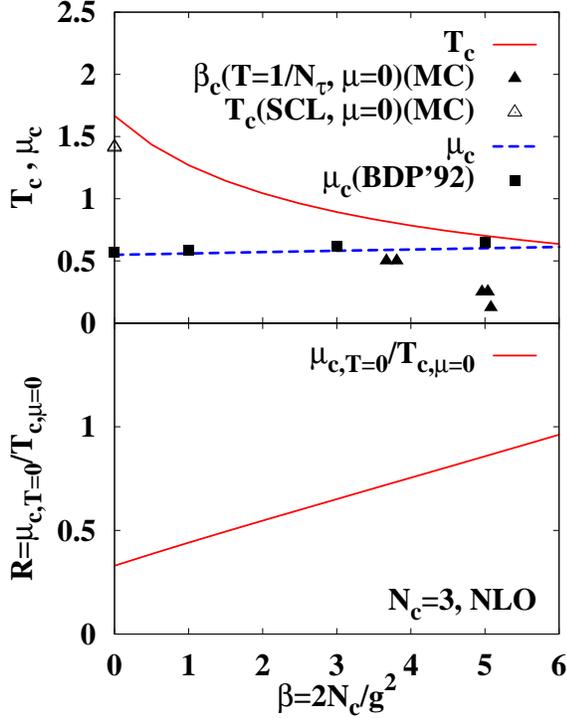}
\caption{Critical temperature, chemical potential, and critical coupling.
In the upper panel,
solid and dashed curves show the NLO results
of $T_{c,\mu=0}$ and $\mu_{c,T=0}$, respectively.
Solid squares show the results of $\mu_{c,T=0}$ in the previous work
by Bilic, Demeterfi and Petersson~\cite{BDP}.
We also show the MC results of the critical coupling ($\beta_c$)
for given $N_\tau=1/T=2, 4~\mathrm{and}~8$ at $\mu=0$ (filled triangles)
and the critical temperature in the strong coupling limit (open triangles).
From the left \protect\cite{deForcrand:2009dh,Boyd:1991fb} 
(the SCL result of $T_c$ with monomer-dimer-polymer simulations),
\protect\cite{Forcrand} ($\beta_c=3.67(2), N_\tau=2$, with a quark mass $m_0=0.025$),
\protect\cite{Forcrand} ($\beta_c=3.81(2), N_\tau=2$, $m_0=0.05$),
\protect\cite{Gottlieb:1987eg}
($\beta_c=4.90(3), 8^3\times 4$ lattice, extrapolated to $m_0=0$),
\protect\cite{D'Elia:2002gd}
($\beta_c=5.037(3), 16^3\times 4$ lattice, $m_0=0.05$),
\protect\cite{Fodor:2001au}
($\beta_c=5.040(2), 6^3\times 4$ lattice, $m_0=0.05$),
and \protect\cite{Gavai:1990ny}
($N_\tau=8$, extrapolated to $m_0=0$).
In the lower panel, we show the ratio $R=\mu_{c,T=0}/T_{c,\mu=0}$.
}\label{Fig:crit}
\end{figure}
%-------------------------------------------------------------------------
% Refs for Nf=4 MC	Lsize etc	mass		Tc and bc
%-------------------------------------------------------------------------
% Boyd:1991fb		MDP,Nt=4	extrapolated	Tc=1.41(3)
% deForcrand:2009dh	MDP,Nt=4	massless	Tc=1.401(2)
% 			MDP,Nt=\infty	massless	Tc=1.417(3)
% Kennedy		\infty^3*2	quench		5.097 \pm 0.001
%			\infty^3*4	quench		5.696 \pm 0.004
%			\infty^3*6	quench		5.877 \pm 0.006
%			\infty^3*8	quench		6.00 \pm 0.02
% Forcrand		Nt=2		0.025		3.67
% Forcrand		Nt=2		0.05		3.81
% Gottlieb:1987eg	8^3*4		0.025		4.96
% Gottlieb:1987eg	8^3*4		extrapolated	4.90 
% D'Elia:2002gd		8^3*4		0.05		5.037(3)
% Fodor:2001au		6^3*4		0.05		5.040(2)
% Gavai:1990ny		Nt=8		extrapolated	5.08 \pm O.08
%-------------------------------------------------------------------------
Along the $T$ axis ($\mu=0$),
the quark number density $\rho_q=\wt$ is always zero,
then the effective potential under the $\wt$ stationary condition
is simply given as $\Feff{eff}(\sigma)=\Feff{eff}(\sigma,\wt=0)$.
In the upper panel of Fig.~\ref{Fig:surf},
we show the effective potential 
at several temperatures ($T/T_c=0, 0.2, \ldots, 1.0, 1.2$) at $\mu=0$
for $\beta=4.5$.
The effective potential has one local minimum in the region $\sigma \geq 0$.
As $T$ becomes large, the minimum point of the effective potential
smoothly decreases to zero from a finite value.
We find that the phase transition along the $T$ axis is the second order
as in the case of SCL~\cite{Fukushima:2003vi,Nishida:2003fb,Kawamoto:2005mq}.

In Fig.~\ref{Fig:crit},
we show the critical temperature at zero chemical potential
$T_{c,\mu=0}$ as a function of $\beta$.
We find that $T_c$ is suppressed as $\beta$ becomes large.
This decrease would be a natural consequence of finite coupling,
since hadrons are less bound than in SCL.
In the present treatment,
the decrease of $T_c$ is caused
by the wave function renormalization factor $Z_\chi$
in Eq.~(\ref{Eq:mq}),
which has a similar effect to the temporal lattice spacing modification.
The second order phase transition temperature
is obtained from the condition $C_2=0$,
where $\Feff{eff}=\sum_n C_n \sigma^n / n!$.
The effective potential $\Feff{eff}(\sigma)$
and auxiliary fields $\Psi=(\varphi_{\tau,s},\omega_{\tau})$ 
are even functions of $\sigma$
in the chiral limit $\partial\Psi/\partial\sigma|_{\sigma\to 0}=0$, 
and the first derivative of auxiliary fields are zero
from the stationary conditions $\partial\Feff{eff}/\partial\Psi=0$.
By using these, we find $C_2$ is given as,
\begin{align}
C_2
\equiv&
\Bigl(\frac{\partial}{\partial\sigma}
+\sum_{\Psi}\frac{\partial\Psi}{\partial\sigma}\frac{\partial}{\partial\Psi}\Bigr)^2\Feff{eff}
\nn\\
\to&\left.\frac{\partial^2\Feff{eff}}{\partial\sigma^2}\right|_{\sigma=0}
=
\left[
\frac{\partial^2\Feff{aux}}{\partial\sigma^2}
+\frac{b_\sigma^2}{Z_\chi^2}\frac{\partial^2\Vq}{\partial \tilde{m}_q^2}
\right]_{\sigma=0}
\nonumber\\
=&\bsig
- \frac{\bsig^2}{Z_\chi^2}\,\frac{N_c(N_c+1)(N_c+2)}
	{3T(N_c+1+2\cosh(N_c\tilde{\mu}/T))}
\ .
\label{Eq:C2}
\end{align}
From the condition $C_2=0$ at $\mu=0$, we find,
\begin{align}
T_{c,\mu=0}=
\frac{T_c^\mathrm{(SCL)}}{Z_\chi^2}
=
\frac{1}{Z_\chi^2}
\frac{d(N_c+1)(N_c+2)}{6(N_c+3)}
\label{eq:Tc}
\ ,
\end{align}
where $T_c^\mathrm{(SCL)}$
represents the second order phase transition temperature 
at $\mu=0$ in the strong coupling limit.
It should be noted that at $\mu=0$ and $\sigma=0$,
$Z_\chi$ does not depend on the auxiliary field.
As shown in Eq.~(\ref{eq:Tc}),
the critical temperature decreases
due to the wave function renormalization factor,
$Z_\chi \geq 1$ at $\sigma=0$.
This mainly originates from
the suppression of the constituent quark mass
$\tilde{m}_{q}=m^{\prime}_q/Z_\chi$.
In this way, the decrease of the critical temperature
is understood
as the $\tilde{m}_q$ modification effects caused by the plaquettes.
The $T_c$ values here are consistent with those in Ref.~\cite{Bilic:1995tq}.

In Fig.~\ref{Fig:crit}, we also show the results of the critical coupling 
in Monte-Carlo simulations with 
$N_\tau=2$~\cite{Forcrand},
$4$~\cite{D'Elia:2002gd,Fodor:2001au,Gottlieb:1987eg}
and $N_{\tau}=8$~\cite{Gavai:1990ny}
temporal lattice sizes.
These results corresponds to 
$T_c=0.5,~0.25~\mathrm{and}~0.125$.
Results with $N_\tau=2$ and $N_\tau=4$
are those with $m_0=0.05, 0.025~\mathrm{or}~0 (\mathrm{extrapolated})$,
and chiral extrapolated results are shown for $N_\tau=8$.
Compared with the results of the critical coupling 
$\beta_c=5.097$ 
in the quenched calculation with $N_\tau=2$~\cite{Kennedy},
$\beta_c$ is significantly smaller with finite masses,
$\beta_c=3.81~\mathrm{and}~3.67$ for 
$m_0=0.05$ and $m_0=0.025$~\cite{Forcrand}.
The monomer-dimer-polymer simulations
on anisotropic lattice~\cite{Boyd:1991fb,deForcrand:2009dh}
give the critical temperature 
$T_c=\gamma_c^2/N_{\tau}=1.401(2)$~\cite{deForcrand:2009dh},
where $\gamma_c$ is the critical anisotropy in the chiral limit.
The decrease of $T_c$ in NLO at finite $\beta$ is not enough 
to explain these MC results,
and higher order effects such as the next-to-next-to-leading order (NNLO)
and Polyakov loop effects would be necessary.
%---------------------------------------------------------------------

%---------------------------------------------------------------------
\subsection{Chiral transition at finite density}\label{subsec:finitemu}
%---------------------------------------------------------------------

At finite $\mu$, 
the quark number density $\rho_q=\wt$ is generally finite
and depends on $\sigma$.
We search for $\wt$ which maximizes $\Feff{eff}(\sigma,\wt)$
for a given $\sigma$,
and we substitute the solution, $\wt=\wt^\mathrm{stat.}(\sigma)$,
in the effective potential.
In the lower panel of Fig.~\ref{Fig:surf},
we show the effective potential 
at several chemical potentials ($\mu/\mu_c=0, 0.2, \ldots, 1.0, 1.2$)
at $T=0$ for $\beta=4.5$.
The effective potential has one local minimum for $\mu$
smaller than the second order critical chemical potential,
$\mu < \mu_{c,T=0}^{(\mathrm{2nd})}$,
and two local minima appear in the larger $\mu$ region.
For $\beta=4.5$, the vacuum
jumps from the NG phase ($\sigma \simeq \sigma_\mathrm{vac.}$)
to the Wigner phase ($\sigma = 0$) at critical chemical potential, $\mu=\mu_c$,
and this transition is the first order.

The first order chiral transition at finite $\mu$
necessarily involve the density gap.
In the case of $\beta=4.5$ and $T=0$ shown
in the lower panel of Fig.~\ref{Fig:sigwt},
the effective potentials at two points in $(\sigma,\wt)$ plane
become equal at $\mu=\mu_c$,
and the first order phase transition takes place.
In the Wigner phase, the quark mass is small (zero in the chiral limit),
then the quark number density is high.
At high densities, the chemical potential effects are reduced
as $\mu \to \tilde{\mu}=\mu-\delta\mu$ as discussed in the previous section.

In the upper panel of Fig.~\ref{Fig:crit},
we show the critical chemical potential at $T=0$, $\mu_{c,T=0}$,
as a function of $\beta$.
In the region of $\beta <6$, the phase transition at $T=0$
is the first order, as in the case of SCL results.
We find that the first order critical chemical potential
$\mu_{c,T=0}^{\mathrm{(1st)}}$
is not largely modified from the strong coupling limit value
$\mu_c^{\mathrm{(SCL,1st)}}\simeq 0.55$.
For example, we find $\mu_c^{\mathrm{(1st)}}\simeq 0.58 (0.60)$ 
at $\beta=3.0 (4.5)$.
This small modification is understood as follows:
In the low temperature region,
the first order phase transition is described 
in terms of the competition between the quark chemical potential
and the constituent quark mass.
Since the temporal plaquette suppresses both,
the relative relations between them are not largely changed.
Hence $\mu_c^{\mathrm{(1st)}}\simeq\mu_c^{\mathrm{(SCL,1st)}}$ follows.
Results by Bilic et al.~\cite{BDP,BKR,Bilic:1995tq} are also shown 
in Fig.~\ref{Fig:crit}.
Our results are qualitatively consistent with their results.

We can now discuss the critical value ratio
$R=\mu_{c,T=0}/T_{c,\mu=0}$,
which characterizes the shape of the phase diagram.
In the lower panel of Fig.~\ref{Fig:crit},
we show this ratio as a function of $\beta$.
As already discussed,
$T_{c,\mu=0}$ rapidly decreases as $\beta$ increases,
while the finite coupling effects give rise to
only small modifications of $\mu_{c,T=0}$.
As a result, the ratio $R$ significantly increases with $\beta$
as shown in the lower panel of Fig.~\ref{Fig:crit}.
The ratio $R$ becomes close to $1$ at $\beta=6$,
and much larger than the SCL results
$R_{\mathrm{SCL}}\sim 0.3-0.45$
\cite{Fukushima:2003vi,Nishida:2003fb,Kawamoto:2005mq}.
The lattice MC results indicate that the critical end point
may locate in the region $\mu/T>1.0$, which suggests $R>1.0$.
Based on the recent MC results ($T_c = 170-200~\mathrm{MeV}$)
and a na\"ive estimate $N_c\mu_c \gtrsim M_N$ ($M_N$ is the nucleon mass),
the expected ratio in the real world would be $R=1.5\sim 3$.
Thus the finite coupling effects are found to increase the ratio $R$
and make it closer to the empirical value.

%---------------------------------------------------------------------
\subsection{Partially chiral restored matter}\label{subsec:PCRM}
%---------------------------------------------------------------------

%---------------------------------------------------------------------
\begin{figure}[tbh]
\Psfig{6cm}{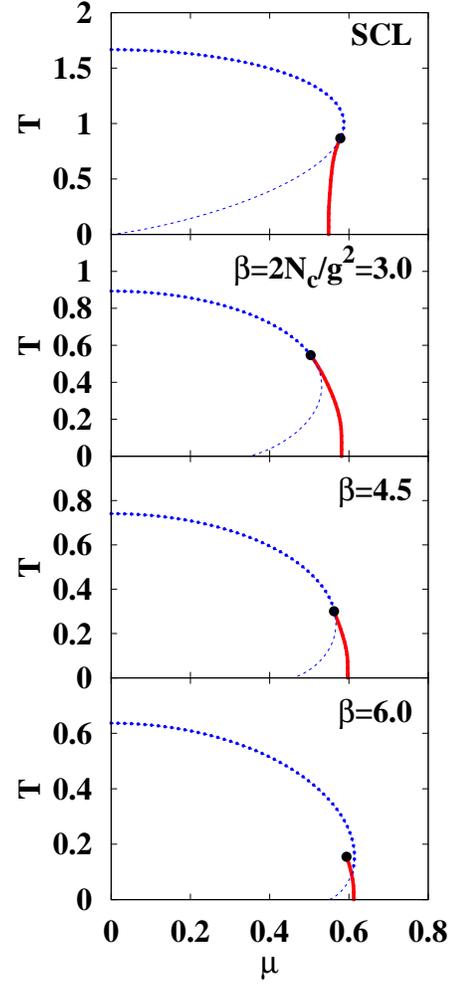}
\caption{(Color online)
The phase diagram for the several value of
$\beta=2N_c/g^2$ in the lattice unit.
The solid and dashed lines
represent the first and second order transition lines,
respectively. The actual transition is described by
the thick dashed and solid lines.
}\label{Fig:PDevol2}
\end{figure}
%---------------------------------------------------------------------

%---------------------------------------------------------------------
\begin{figure}[tbh]
\Psfig{8.0cm}{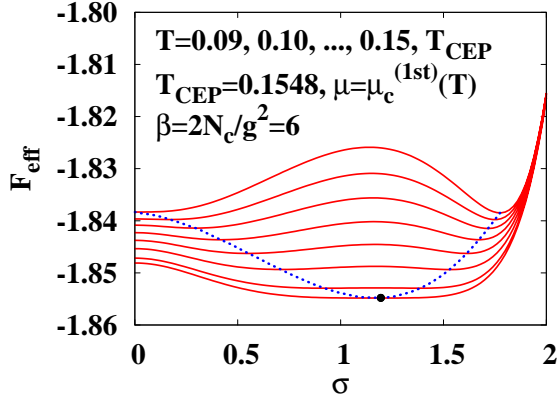}
\caption{
Effective potential on the first phase transition boundary
at $\beta=6$. Solid lines show the results at $T=0.12, 0.13, 0.14, 0.15$
and $T_\mathrm{CEP}$, where $T_\mathrm{CEP}=0.1548$,
and dashed lines connect the coexisting equilibrium points on the boundary.
}\label{Fig:surf60CEP}
\end{figure}
%---------------------------------------------------------------------

One of the characteristic features of the present treatment is that
the second order critical chemical potential $\mu_c^{(\mathrm{2nd})}$ 
is finite even at $T=0$.
The second order critical chemical potential $\mu_c^{(\mathrm{2nd})}$ 
for a given $T$ is obtained by solving the condition $C_2=0$
at finite chemical potential.
By using Eq.~(\ref{Eq:C2}),
the shifted second order critical chemical potential is obtained as,
\begin{align}
\tilde{\mu}_c^{(\mathrm{2nd})}
=\frac{T}{N_c}
\mathrm{arccosh}
\biggl[1+\frac{(N_c+3)(T_c^\mathrm{(SCL)}/Z_\chi^2-T)}{2T}\biggr]
\ .
\end{align}
In rhs, $\sigma=0$ is assumed.
This equation is an implicit equation, which should be solved with
the condition $\wt=\rho_q$ simultaneously;
$Z_\chi$ in rhs is a function of $\wt$,
which is a function of $\tilde{\mu}$.
The second order critical chemical potential differs
from $\tilde{\mu}_c^{(\mathrm{2nd})}$ by $\delta\mu$,
\begin{align}
\mu_c^{\mathrm{(\mathrm{2nd})}}
&={\tilde{\mu}}_c^{(\mathrm{2nd})} +\delta\mu
={\tilde{\mu}}_c^{(\mathrm{2nd})}+\log\sqrt{Z_+/Z_-}
\label{Eq:amuc2nd}\ .
\end{align}
At $T=0$,
$\tilde{\mu}_c^{(\mathrm{2nd})}$ becomes zero.
In SCL, 
we do not have the second term in Eq.~(\ref{Eq:amuc2nd})
and $\mu_c^{\mathrm{(2nd)}}$ approaches zero at small $T$.
In NLO,
the second term in Eq.~(\ref{Eq:amuc2nd}) is finite
at finite $\beta$.
As a result, there is a possibility that $\mu_c^{(\mathrm{2nd})}$ overtakes
$\mu_c^{(\mathrm{1st})}$, which leads to the appearance
of the partially chiral restored matter~\cite{QY,Lat08_Miura}.

We numerically find that
$\mu_c^\mathrm{(2nd)}$ can overtake $\mu_c^\mathrm{(1st)}$
in the case $\beta \gtrsim 4.5$.
We show the phase diagrams with $\beta=3.0, 4.5~\mathrm{and}~6.0$ 
in comparison with the SCL phase diagram in Fig.~\ref{Fig:PDevol2}.
The tri-critical point (TCP) starts to deviate from
the second order phase transition boundary at $\beta\sim 4.5$,
and becomes a critical end point (CEP) at larger $\beta$
even in the chiral limit.
When the CEP exists off the second order phase transition boundary,
we have the temperature region,
where the second order critical chemical potential is larger than
the first order one, $\mu_c^\mathrm{(2nd)}(T)>\mu_c^\mathrm{(1st)}(T)$.
In this temperature region, $\Feff{eff}$ at $\mu=\mu_c^\mathrm{(1st)}$
should have two local minima in the $\sigma>0$ region 
as shown in Fig.~\ref{Fig:surf60CEP}:
At $\mu=\mu_c^\mathrm{(1st)}$, $\Feff{eff}$ at two local minima are equal,
and the local minimum with smaller $\sigma$ cannot be at $\sigma=0$,
since the curvature of $\Feff{eff}$ is negative at around $\sigma=0$
(i.e. $C_2<0$) in the chemical potential region $\mu<\mu_c^\mathrm{(2nd)}$.

In the temperature region
where the condition $\mu_c^{\mathrm{(2nd)}}>\mu_c^{\mathrm{(1st)}}$
is satisfied, we have three chemical potential regions,
$\mu < \mu_c^{\mathrm{(1st)}}$,
$\mu_c^{\mathrm{(1st)}}< \mu < \mu_c^{\mathrm{(2nd)}}$,
and $\mu > \mu_c^{\mathrm{(2nd)}}$.
The vacuum is in the NG phase in the first region,
where the chiral condensate is large enough.
In the third region, the chiral condensate is completely zero,
and it is in the Wigner phase.
In the second region, the chiral symmetry is weakly but spontaneously broken,
and a partially chiral restored (PCR) matter is realized~\cite{QY,Lat08_Miura}.
It is interesting to investigate the transitions among them.
As $\mu$ increases, the $\sigma$ jumps
from the NG local minimum to the PCR local minimum with $\sigma>0$,
as we can guess from the $\Feff{eff}$ behavior in Fig.~\ref{Fig:surf60CEP}.
For larger $\mu$, the chiral condensate in PCR matter decreases,
and the Wigner phase ($\sigma=0$) is realized at $\mu=\mu_c^\mathrm{(2nd)}$.
%

%===============================================================
\begin{table*}[tb]
\caption{The truncation schemes in NLO-A, B, C and D.
In NLO-C and D,
${\cal O}(1/g^4)$ terms in $\tilde{m}_q$ are truncated to be
$\tilde{m}_q^\mathrm{(NLO-C,D)}
=(b_\sigma\sigma+m_0)(1-\beta_\tau\varphi_\tau)+2\beta_s\varphi_s\sigma$.
}\label{Tab:NLO_ABC}
\begin{center}
\begingroup
\renewcommand{\arraystretch}{2.5}
\begin{tabular}{c|c|c|c|c}
\hline\hline
&$\delta\mu$
&$\tilde{m}_q$
&\Disp{\Delta\Feff{aux}}
&\Disp{\Vq}
\\\hline\hline
NLO-A&\Disp{\log\sqrt{\frac{Z_+}{Z_-}}}
     &\Disp{\frac{m_q}{\sqrt{Z_+Z_-}}}
     &\Disp{-N_c\log\sqrt{Z_+Z_-}}
     &\Disp{\Vq(\tilde{m}_q,\tilde{\mu},T)}
\\\hline
NLO-B&\Disp{\beta_{\tau}\omega_{\tau}}
     &\Disp{\frac{m_q}{1+\beta_{\tau}\varphi_{\tau}}}
     &\Disp{-N_c\log(1+\beta_{\tau}\varphi_{\tau})}
     &\Disp{\Vq(\tilde{m}_q,\tilde{\mu},T)}
\\\hline
NLO-C&\Disp{\beta_{\tau}\omega_{\tau}}
     &\Disp{\tilde{m}_q^\mathrm{(NLO-C)}}
     &\Disp{-N_c\beta_{\tau}\varphi_{\tau}}
     &\Disp{\Vq(\tilde{m}_q,\tilde{\mu},T)}
\\\hline
NLO-D&$0$
     &\Disp{\tilde{m}_q^\mathrm{(NLO-D)}}
     &\Disp{-N_c\beta_{\tau}\varphi_{\tau}}
     &\Disp{\Vq(\tilde{m}_q,\mu,T)
	-\beta_\tau\wt\frac{\partial\Vq}{\partial\mu}}
\\\hline\hline
\end{tabular}
\endgroup
\end{center}
\end{table*}
%===============================================================

The appearance of PCR matter, or equivalently,
$\mu_c^{\mathrm{(1st)}}< \mu < \mu_c^{\mathrm{(2nd)}}$ region
may stem from the multi-order parameter treatment~\cite{QY,Lat08_Miura}.
To clarify this point, we examine several truncation schemes,
where the effective potential $\Feff{eff}(\sigma,\rho_q)$
systematically reduces to that with a single order parameter $\sigma$.
And we would check the disappearance of the PCR matter.

The first treatment is the same as that we have discussed
in previous subsections, and abbreviated as NLO-A.
In the second treatment (NLO-B),
${\cal O}(1/g^4)$ contributions in $Z_\chi$ and $\tilde{\mu}$ are truncated as,
\begin{align}
Z_\chi^\mathrm{(NLO-B)}
	=&1+\beta_\tau\varphi_\tau
\ ,\label{Eq:NLO-B-Z}\\
\tilde{\mu}^\mathrm{(NLO-B)}
	=&\mu-\beta_\tau\omega_\tau
\ .\label{Eq:NLO-B-mu}
\end{align}
In this treatment, we find that $\varphi_\tau$ and $\omega_\tau$
couple to quarks separately
through $\tilde{m}_q$ and $\tilde{\mu}$, respectively.
In the third prescription (NLO-C),
we further truncate ${\cal O}(1/g^4)$ terms in 
$\tilde{m}_q$ and in $\log Z_\chi$.
\begin{align}
\tilde{m}_q^\mathrm{(NLO-C)}
=&(b_\sigma\sigma+m_0)(1-\beta_\tau\varphi_\tau)+2\beta_s\varphi_s\sigma
\ ,\label{Eq:NLO-C-mq}\\
\Delta \Feff{aux}\equiv& -N_c\log Z_\chi
\nn\\
\approx& -N_c\beta_\tau\varphi_\tau
\quad\mathrm{(NLO-C)}
\ .
\end{align}
It is also possible to expand $\Vq$ 
with respect to $\delta\mu=\mu-\tilde{\mu}$ (NLO-D),
\begin{align}
\Vq^\mathrm{(NLO-D)}(\tilde{m}_q;\tilde{\mu},T)
\simeq \Vq(\tilde{m}_q;\mu,T)
-\beta_\tau\omega_\tau\frac{\partial\Vq}{\partial\mu}
\ .
\end{align}
These truncation schemes are summarized in Table~\ref{Tab:NLO_ABC}.

Stationary conditions in NLO-B, C and D are solved 
in a similar way to NLO(NLO-A).
In NLO-B and C,
$\wt$ is still an implicit
function of $\sigma$,
$\omega_{\tau}=\rho_q(\sigma,\omega_\tau;\mu,T)$,
and the multi-order parameter property is still kept.
In NLO-D, $\omega_\tau$ is explicitly obtained as a function of $\sigma$,
\begin{align}
\wt^\mathrm{(NLO-D)}=-\frac{\partial\Vq(m_q(\sigma);\mu,T)}{\partial\mu}
\ ,
\end{align}
where the rhs does not contain $\omega_\tau$.
The dynamics is described by a single order parameter $\sigma$.
In this meaning, the NLO-D gives a similar formulation
to those in the previous works~\cite{BDP,BKR,Bilic:1995tq}.

%----------------------------------------------------------------------*
\begin{figure}[bt]
\Psfig{8.0cm}{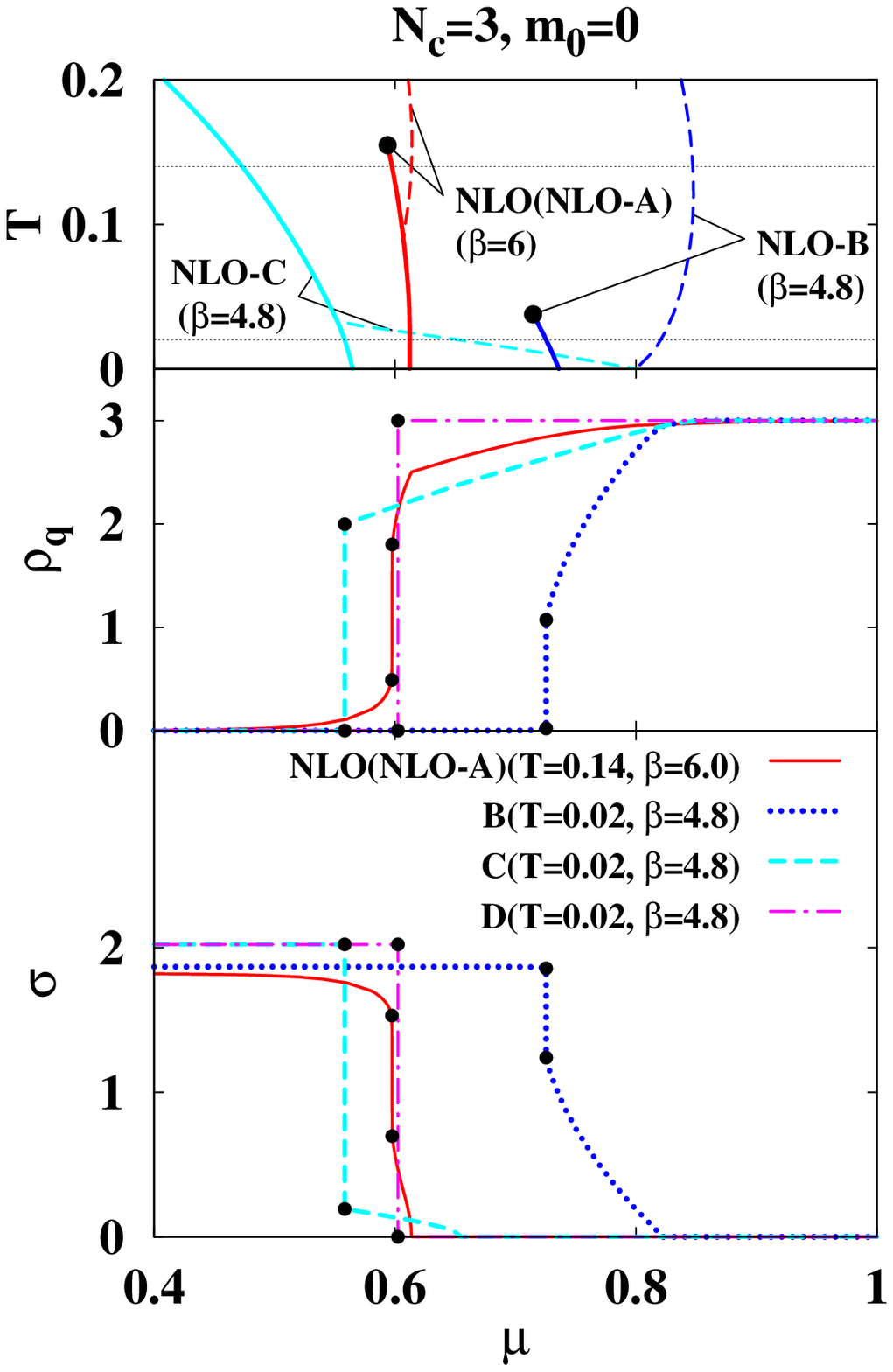}
\caption{
In the upper panel,
solid and dashed curves show the first and second order
phase transition boundaries,
and dots show the critical end point.
In the middle and lower panels,
solid, dotted, dashed, and dot-dashed curves
show the results of NLO-A, B, C and D, respectively,
and dots and open squares show the first order transition points.
}\label{Fig:pdrhosig}
\end{figure}
%----------------------------------------------------------------------*

In the upper panel of Fig.~\ref{Fig:pdrhosig},
we show the phase diagrams in NLO-A, B and C in the large $\mu$ region.
In NLO-A and B,
the maximum temperature of the first order phase boundary decreases,
and the critical point deviates from the second order phase transition boundary 
at $\beta \simeq 4.5$ and $3.0$ in NLO-A and NLO-B, respectively.
In NLO-C, the second order critical chemical potential
$\mu_c^{(\mathrm{2nd})}$ at $T=0$
overtakes the first order one at $\beta \simeq 3.5$.
Between the first and second order phase boundaries, we find PCR matter.
In NLO-D, PCR matter does not appear in any region
of $(T,\mu,\beta)$.

In the middle and lower panels of Fig.~\ref{Fig:pdrhosig}, 
we show the comparison of $\rho_q$ and $\sigma$
in the present treatments NLO-A, B, C and D\cite{QY,Lat08_Miura}.
The gradual increase of the quark number density $\rho_q$
after the first order transition is a common feature
of the multi-order parameter treatments~\cite{QY,Lat08_Miura}.
This means that $\rho_q$ is high, but still smaller than
the maximum density $N_c$ in PCR matter at small $T$.
At low temperatures,
we can investigate the appearance of the PCR matter
more intuitively.
The quark number density 
$\rho_q=-\partial\Feff{eff}/\partial\mu$ is evaluated as,
\begin{align}
\frac{\rho_q}{N_c}=\frac{2\sinh\bigl[N_c\tilde{\mu}/T\bigr]}
{X_{N_c}+2\cosh\bigl[N_c\tilde{\mu}/T\bigr]}
\ 
\raisebox{-2ex}{$\stackrel{\longrightarrow}{{\scriptstyle T\to 0}}$}
\ 
\frac{x^{N_c}}{1+x^{N_c}}
\ ,\label{Eq:rho}
\end{align}
where $x=\exp[-(E_q-\tilde{\mu})/T]$.
When $E_q>\tilde{\mu}$ is satisfied at small $T$,
we obtain $x\to 0$ and $\rho_q\to 0$,
while $E_q<\tilde{\mu}$ leads to $x\to\infty$ and $\rho_q\to N_c$.
Medium density  $0<\rho_q<N_c$ can appear 
only in the case where the energy and chemical potential balances,
$E_q=\tilde{\mu}$, and 
$x$ stays finite at $T=0$.
Since $\tilde{\mu}$ is a decreasing function of $\omega_\tau$,
we may have a medium density solution of Eq.~(\ref{Eq:rho}) in the region
$\tilde{\mu}(\sigma,\omega_\tau=N_c)<E_q(\sigma,\omega_\tau)<\mu$.
Specifically in NLO-B and C,
$E_q=\tilde{\mu}$ is found to be equivalent
to the density condition $\rho_q=(\mu-E_q)/\beta_{\tau}$,
which can take the a medium value.
In the large $\beta$ region,
this medium density matter can emerge in equilibrium
and corresponds to the PCR matter
as indicated in Fig.~\ref{Fig:pdrhosig}.
Also in NLO-A, PCR matter appears in a similar mechanism
at finite $T$.
Thus the multi-order parameter treatment 
is essential to obtain the PCR matter at low $T$,
and we observe the two chiral transitions as $\mu$ increases.

Now we have found following common properties
as long as the quark number density is treated as the order parameter
in addition to the chiral condensate~\cite{QY,Lat08_Miura}.
(I)~The partially chiral restored (PCR) matter
can appear in the large $\beta$ region,
(II)~PCR sits next to the hadronic Nambu-Goldstone (NG) phase
in the larger $\mu$ direction,
(III)~the quark number density is high as $\Od(N_c)$ in PCR,
(IV)~in PCR matter, the effective chemical potential
is adjusted to the quark excitation energy,
and
(V)~the second order chiral transition 
to the Wigner phase follows
after NG$\to$PCR transition.
All these properties would be the essence of
the quarkyonic matter and transition proposed 
in Ref.~\cite{QY_McLerran}.
In the previous work, 
the quark-driven Polyakov loop evaluated in SC-LQCD is shown
to be small as ${\cal O}(1/N_c)$~\cite{Faldt1986},
and it would not grow much at low temperatures.
This feature is also consistent with 
the proposed property of the quarkyonic matter.

The quarkyonic matter is originally defined as the
confined high density matter 
at large $N_c$~\cite{QY_McLerran},
and recently investigated by using the PNJL model~\cite{QY_NJL}.
In order to discuss the deconfinement dynamics,
the higher order of $1/g^2$ expansion would be essential,
and a subject to be studied in future.

%---------------------------------------------------------------------
\subsection{Phase diagram evolution}\label{subsec:PDevol}
%---------------------------------------------------------------------

%---------------------------------------------------------------------
\begin{figure}[tb]
\Psfig{8.0cm}{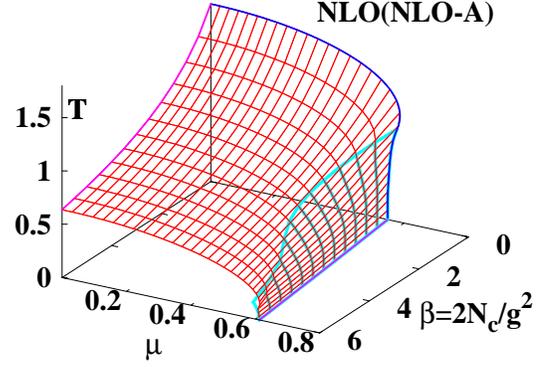}
\caption{
The phase diagram evolution 
with the finite coupling effect $\beta=2N_c/g^2$ in the lattice unit.
Thin line surface shows the boundaries
between the chiral broken and restored phases,
and thick line surface shows the first order boundaries.
}\label{Fig:PDevol}
\end{figure}
%-----------------------------------------------------------------------

%---------------------------------------------------------------------
\begin{figure}[tb]
\Psfig{8.0cm}{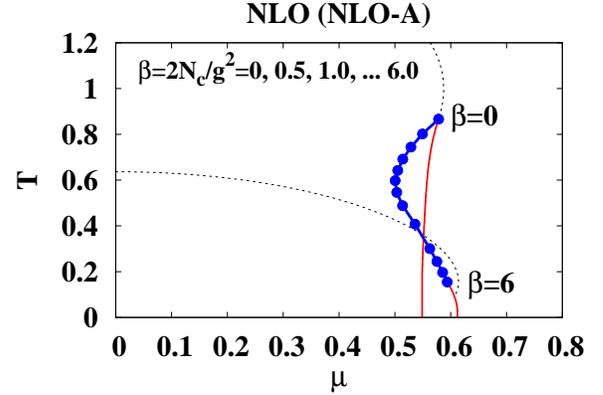}
\caption{
Coupling dependence of the critical point.
Filled circles show the critical points,
and solid and dotted lines show the first and second order
boundaries for $\beta=0$ and $6$.
}\label{Fig:CEPevol}
\end{figure}
%-----------------------------------------------------------------------

We shall now discuss the phase diagram
evolution with $\beta$.
In Fig.~\ref{Fig:PDevol},
we show dependence of the phase diagram on $\beta$
in NLO(NLO-A).
As $\beta$ increases,
the second order phase boundary is compressed in the temporal direction
according to the decrease of $T_c^{(\mathrm{2nd})}$.
Phase transition of cold ($T=0$) dense matter is calculated to be
the first order,
and the critical chemical potential does not move much.

Thick lines in Fig.~\ref{Fig:PDevol} show the coupling dependence of
the first order phase transition boundaries.
As more clearly seen in Fig.~\ref{Fig:PDevol2},
the slope of the first order phase transition boundary
becomes negative at finite coupling,
and it is natural from the Clausius-Clapeyron relation.
We expect that the entropy and quark number density is higher
in the Wigner phase, 
$\Delta s=s^\mathrm{(W)}-s^\mathrm{(NG)}>0$
and $\Delta\rho_q=\rho_q^\mathrm{(W)}-\rho_q^\mathrm{(NG)}>0$,
where $s^\mathrm{(W,NG)}$ and $\rho_q^\mathrm{(W,NG)}$ denote
the entropy and quark number density, respectively,
in the Wigner and the NG phases.
With this expectation, the slope of the first order phase boundary
from the Clausius-Clapeyron relation, $d\mu/dT=-\Delta s/\Delta\rho_q$,
should be negative.
This improvement from SCL
may be related to the gradual increase of the quark number
density in the PCR matter and in the Wigner phase.
In SCL, the quark number density jumps to an almost saturated value,
$\rho_q \sim N_c$, at low $T$ at $\mu=\mu_c^{(1st)}$.
In this case, the lattice sites are almost filled by quarks,
and the entropy density will be very small in the Wigner phase.
This density saturation is a lattice artifact,
and is expected to be weaken at finite $\beta$.
With NLO effects,
the vector field $\wt$ suppresses the sudden increase of $\rho_q$,
and the quark number density gradually increases
after the first order phase transition.
We have discussed this feature in the PCR matter,
and it also applies to the quark matter in the Wigner phase.

We also find that the slope $d\mu/dT$
is always negative in whole $T-\mu$ plane at finite $\beta$.
This point is different from the previous NLO works~\cite{Bilic:1995tq}.

The end point of the first order phase transition boundary
is the critical point,
which is either the tri-critical point (TCP) or the critical end point (CEP).
As we already discussed in the previous subsection,
the TCP at small $\beta$ deviates from the second order phase transition line
at $\beta \gtrsim 4.5$, and becomes the CEP.
The temperature of this critical point gradually decreases,
while the chemical potential stays in a narrow range as $\beta$ increases.
We show the evolution of the critical point with $\beta$
in Fig.~\ref{Fig:CEPevol}.
The decrease of the critical point temperature, $T_{CP}$,
is consistent with the results in the NJL and PNJL models~\cite{NJL-Vector}.
In these works, it is demonstrated that $T_{CP}$ decreases
as we adopt a larger vector coupling relative to the scalar coupling.
In the present work, $\beta_\tau\wt$ is regarded as the vector potential
for quarks, and it grows as $\beta$ increases.

In the continuum limit,
one species staggered QCD would become
the four flavor QCD with degenerate masses~\cite{stagg},
where the chiral transition is expected to be
the first order due to anomaly contributions~\cite{Pisarski:1983ms}.
The present behavior of the critical point shows
that the NLO SC-LQCD does not contain anomaly effects.
%---------------------------------------------------------------------
\section{Concluding Remarks}\label{sec:CR}
%---------------------------------------------------------------------

We have investigated the chiral phase transition
in the strong coupling lattice QCD
at finite temperature ($T$) and chemical potential ($\mu$)
with finite coupling ($\beta=2N_c/g^2$) effects.
We have derived an analytic expression of the effective potential
using one species of staggered fermion
in the leading (strong coupling limit; SCL) and next-to-leading order (NLO)
of the strong coupling ($1/g^2$) expansion
and in the leading order of the large dimensional ($1/d$) expansion.
We have focused our attention on
the phase diagram evolution.

From the NLO effective action, 
we have derived the effective potential
under the mean field approximation
based on a self-consistent treatment of NLO effects
with a recently proposed 
extended Hubbard-Stratonovich (EHS) transformation~\cite{QY,Lat08_Miura}.
Then the quark number density ($\rho_q$) is naturally introduced
as an order parameter.
NLO contributions are expressed via
the shift of the constituent quark mass,
dynamical chemical potential and 
the quark wave function renormalization factor.
The NLO effective potential is found to become a function of
$T$, $\mu$, $\beta$,
the chiral condensate $\sigma$
and quark number density $\rho_q$.
Such a formulation has been essential
in order to investigate the mechanism of
the phase diagram evolution with $\beta$.
The phase diagram has been obtained by performing the
minimum search of the effective potential
in the multi-order parameter treatment.

The effective constituent quark mass $\tilde{m}_{q}$
is found to be suppressed as $\beta$ increases.
As a result, the critical temperature $T_c$ decreases
and becomes closer to the Monte-Carlo results 
at $\mu=0$~\cite{Forcrand,Gottlieb:1987eg,Gavai:1990ny},
while it is still larger than the MC data.
The effective quark chemical potential $\tilde{\mu}$
is also suppressed as $\beta$ becomes larger.
We have found the small modification in the critical chemical potential $\mu_c$
at low $T$.
In this way, the ratio $R=\mu_{c,T=0}/T_{c,\mu=0}$ becomes larger
and closer to the empirical value.
The $\beta$ dependences of $T_{c,\mu=0}$ and $\mu_{c,T=0}$
are consistent with the previous results~\cite{Bilic:1995tq}.
The first order phase boundary is found to satisfy $d\mu/dT \leq 0$
at finite $\beta$.
This behavior is natural from the Clausius-Clapeyron relation,
and is different from the SCL results and previous results with NLO
effects~\cite{Bilic:1995tq}.
In the phase diagram evolution,
the tri-critical point is found to move in the lower $T$ direction.
This trend is consistent with
model results~\cite{NJL-Vector}.
Partially chiral restored (PCR) matter is found to appear
in the low $T$ and the large $\mu$ region
with $\beta \gtrsim 4.5$.
We have shown that the multi-order parameter ($\sigma,\wt$)
treatment is essential in describing PCR matter,
where the effective chemical potential is automatically adjusted
to the quark excitation energy.

We have discussed the NLO results in the region $\beta\leq6$,
expecting that the strong coupling expansion is convergent
even in the region $\beta=5\sim6$.
In the pure Yang-Mills theory, 
the character and strong coupling expansions
seem to be convergent in the region
of $2N_c/g^2\simeq 2N_c$ for color SU(2)~\cite{Munster:1980iv} 
and SU(3)~\cite{Drouffe:1983fv}.
For color SU(3), the MC simulations indicate that the critical coupling
$\beta_c$ at $\mu=0$ seems to be a smooth function of
$T=1/N_\tau$~\cite{deForcrand:2009dh,Boyd:1991fb,Forcrand,Fodor:2001au,D'Elia:2002gd,Gottlieb:1987eg,Gavai:1990ny}
and reaches $\beta=5.08$ for $N_\tau=8$~\cite{Gavai:1990ny}.
When we take into account 
the next-to-next-to-leading order (NNLO) contributions
in SC-LQCD with quarks for color SU(3),
$T_{c,\mu=0}$ and $\mu_{c,T=0}$ are found to be very similar to those
in NLO in the region $\beta\leq 6$~\cite{NNLO}.
These observations suggest that
the strong coupling expansion does not break down in the region $\beta\leq 6$.
It would be necessary to investigate the NNLO effects on the critical point
and PCR matter in order to examine the present results.

There are several points to be discussed further.
When we take into account NNLO contributions,
the Polyakov loop can appear from two plaquettes.
Hence it becomes possible to investigate
the phase transitions with three order parameters,
$\sigma$, $\rho_q$ and the Polyakov loop.
In addition, the Polyakov loop contributions in the NNLO
may modify the $\beta$ dependence of $T_c$.
The higher order of the $1/d$ expansion
is also an important subject to be studied.
The baryonic contributions are included in the sub-leading order
of the $1/d$ expansion,
and would be essential to solve a challenging problem:
nuclear matter on the lattice.

%%%%%%%%%%%%%%%%%%%%%%%%%%%%%%%%%%%%%%%%%%%%%%%%%%%%%%%%%%%%%%%%%%%%%%%%
\section*{Acknowledgments}
We would like to thank
Philippe de Forcrand, Koichi Yazaki, Koji Hashimoto,
for useful discussions.
This work is supported in part by KAKENHI,
under the grant numbers,
17070002		% (Priority area)
and 
19540252,		% ((C)(2), 2003), Ohnishi
the Global COE Program
"The Next Generation of Physics, Spun from Universality and Emergence",
and the Yukawa International Program for Quark-hadron Sciences (YIPQS).

%%%%%%%%%%%%%%%%%%%%%%%%%%%%%%%%%%%%%%

%%%%%%%%%%%%%%%%%%%%%%%%%%%%%%%%%%%%%%
\end{document}